\documentclass{aa}

\usepackage{epsfig,graphicx,natbib}
\usepackage{longtable}
\usepackage{amssymb}
\usepackage{amsmath}
\usepackage{mathrsfs} 
\usepackage{color}
\usepackage{subcaption}
\usepackage{booktabs}
\usepackage{xspace}
\usepackage{float}
\usepackage{longtable}
\usepackage{breqn}
\usepackage{algpseudocode}
\usepackage{algorithm}
\usepackage{hyperref}
\usepackage[outercaption]{sidecap}  

\usepackage{todonotes}

\newcommand{\RR}{\mathbb{R}}

\newcommand{\sgra}{Sgr~A$^*$\xspace}

\defcitealias{Issaoun2022}{Issaoun, Wielgus et al. 2022}
\defcitealias{Jorstad2022}{Jorstad, Wielgus et al. 2022}
\defcitealias{Mueller2023}{Paper I}
\defcitealias{Zhang2008}{Zhang \& Li (2007)}
\defcitealias{Pardalos2017}{Pardalos et al. 2017}

\begin{document}

\title{Using multiobjective optimization to reconstruct interferometric data (II): polarimetry and time dynamics}

\author{Alejandro Mus \inst{1,2}
\and Hendrik M\"uller \inst{3}\thanks{Both first authors have contributed equally to this work.}
\and Ivan Martí-Vidal \inst{1,2}
\and Andrei Lobanov \inst{3}
}

\institute{
  Departament d’Astronomia i Astrof\'isica, Universitat de Val\`encia, C. Dr. Moliner 50, 46100 Burjassot ,Val\`encia, Spain \\
  \email{alejandro.mus@uv.es}
  \and
  Observatori Astron\`omic, Universitat de Val\`encia, Parc Cient\'ific, C. Catedr\`atico Jos\'e Beltr\'an 2, 46980 Paterna, Val\`encia, Spain
  \and 
   Max-Planck-Institut für Radioastronomie, Auf dem Hügel 69, D-53121 Bonn (Endenich), Germany \\ \email{hmueller@mpifr-bonn.mpg.de}
}

\date {Received  / Accepted}

\authorrunning{Mus+M\"uller}
\titlerunning{Using multiobjective optimization to reconstruct interferometric data (II)}

\abstract
%context heading (optional)
{In Very Long Baseline Interferometry (VLBI), signals from multiple antennas combine to create a sparsely sampled virtual aperture, its effective diameter determined by the largest antenna separation. The inherent sparsity makes VLBI imaging an ill-posed inverse problem, prompting the use of algorithms like the Multiobjective Evolutionary Algorithm by Decomposition (MOEA/D), as proposed in the first paper of this series.}
% aims heading (mandatory)
{This study focuses on extending MOEA/D to polarimetric and time dynamic reconstructions, particularly relevant for the VLBI community and the Event Horizon Telescope Collaboration (EHTC). MOEA/D's success in providing a unique, fast, and largely unsupervised representation of image structure serves as the basis for exploring these extensions.}
% methods heading (mandatory)
{The extension involves incorporating penalty terms specific to total intensity imaging, time-variable, and polarimetric variants within MOEA/D's multiobjective, evolutionary framework. The Pareto front, representing non-dominated solutions, is computed, revealing clusters of proximities. Testing MOEA/D with synthetic datasets representative of EHTC's main targets demonstrates successful recovery of polarimetric and time-dynamic signatures despite sparsity and realistic data corruptions.}
% results heading (mandatory)
{MOEA/D's extension proves effective in the anticipated EHTC setting, offering an alternative and independent claim to existing methods. It not only explores the problem globally but also eliminates the need for parameter surveys, distinguishing it from Regularized Maximum Likelihood (RML) methods. MOEA/D emerges as a novel and useful tool for robustly characterizing polarimetric and dynamic signatures in VLBI datasets with minimal user-based choices.}
% conclusions heading (optional)
{Future work aims to address the last remaining limitation of MOEA/D, specifically regarding the number of pixels and numerical performance, to firmly establish it within the VLBI data reduction pipeline.}

\maketitle

\section{Introduction}

For Very Long Baseline Interferometry (VLBI) multiple antennas in a radio-interferometric array are combined to achieve unmatched astronomical resolutions. The source is simultaneously observed by an array of radio antennas. As described by the van-Cittert-Zernike theorem~\citep{Thompson2017}, the correlated products of the signals recorded by pairs of antennas are approximately the Fourier transform of the true sky brightness distribution with a Fourier frequency determined by the baseline separating the two antennas on the sky plane.

Image reconstruction deals with the problem of recovering the sky brightness distribution from these measurements. With a full sampling of the Fourier domain (commonly referred to as the $u,v$ plane), the original image could be retrieved by an inverse Fourier transform. However, due to the limited number of antennas and the limited amount of observing time, only a sparse subset of all Fourier coefficients is measured (the subset of all observed spatial Fourier frequencies is commonly referred to as the $u,v$-coverage). Moreover, the observations are corrupted by additional thermal noise and instrumental calibration effects. Hence, the imaging problem is an ill-posed inverse problem.

Particularly global VLBI at mm-wavelengths pose a number of additional challenges: The number of antennas is small, the visibilities are less well calibrated (in fact typically the phases are poorly constrained) and the signal-to-noise ratio is worse compared to denser arrays operating at longer wavelengths. All in all, the reconstruction problem is under-constrained and, due to missing data and the need for self-calibration, nonconvex and {possibly} multimodal. Thus, the imaging relies on strong prior information, {and a different selection of the regularization terms or prior distributions may produce significantly different image features.}

CLEAN and its variants \citep{Hogbom1974, Clark1980, Bhatnagar2004, Cornwell2008, Rau2011} are the de-facto standard method for imaging since decades. However, recent years saw the ongoing development of novel imaging algorithms for the global VLBI data regime, particularly inspired by the needs of the Event Horizon Telescope (EHT), in three main families: super-resolving CLEAN-based algorithms \citep{Mueller2022b}, Bayesian methods \citep{Arras2021, Broderick2020, Tiede2022}, and regularized maximum likelihood (RML) methods \citep{Homma14, Chael2016, Chael2018, Akiyama2017, Akiyama2017b, Mueller2022}. These methods have been proven successful on synthetic data \citep{eht2019d, eht2022c} and in a wide range of frontline observations with the EHT \citep{eht2019d, Kim2020, Janssen2021, eht2022c, Issaoun2022, Jorstad2023}, the Global Millimetre VLBI Array (GMVA) observations \citep{Zhao2022}, as well as with space-VLBI \citep{Fuentes2022, Mueller2022c, Kim2023}. The issue {that the image structure is only weakly constrained by the data, and multiple solutions may fit the observed data,} is handled in these frameworks either by manual interaction in CLEAN (CLEAN windows, hybrid imaging and self-calibration), a global posterior evaluation for Bayesian methods, or the combination of various data terms and regularization terms for RML methods. In particular, to overcome the limitation imposed by the sparse uv-coverage and challenging calibration when observing at $230\,\text{GHz}$, the EHT performed extensive and successful surveys exploring different parameter combinations. These were utilized to select a top-set of hyper-parameters, but also served the purpose of a synthetic data verification with a variety of source morphologies to build confidence in the reconstructed images (in particular for RML methods)~\citep[see][]{eht2019d,eht2022c}. In this work, we focus on improving the top-set selection strategy.

Recently, we proposed method, MOEA/D, that presents an alternative claim \citepalias{Mueller2023}. MOEA/D approaches the image reconstruction by a multiobjective formulation similar in philosophy to parameter surveys that are common for RML methods \citepalias{Mueller2023, Zhang2008}. The nondominated, locally optimal solutions to the image reconstruction problem are identified by Pareto optimality. In this way, MOEA/D directly fits into the framework of a multimodal, nonconvex imaging problem, since it investigates all clusters of optimal features simultaneously \citep{Zhang2008}. As a particular benefit of MOEA/D, it is relatively fast and largely unsupervised. Compared to CLEAN, no human supervision is needed anymore. Compared to Bayesian methods {with manually chosen prior distributions} and RML methods, MOEA/D has a small number of hyperparameters, explicitly explores {inherent degeneracies} and may not require parameter surveys \citepalias{Mueller2023}. {An alternative strategy (to mapping the impact of various regularization assumptions by the Pareto front, and thus reducing the number of critical free hyperparameters) would be to infer the priors or regularization terms together with image structure as realized by \texttt{resolve} in a Bayesian framework \citep{Arras2019}, or by particle swarm optimization in an RML setting \citep{Mus2023b}.}

The VLBI community shows great interest in two extensions of the classical imaging problem: Polarimetric reconstructions \citep[e.g.][]{Lister2018, Kramer2021, eht2021a, eht2021b, Ricci2022, Johnson2023} and time dynamic reconstructions \citep[e.g.][]{Johnson2017, Bouman2017, Lister2018, Lister2021, Satapathy2022, Wielgus2022, Farah2022, Johnson2023}. The option to recover polarimetric movies is of particular importance for the EHT and its possible successors \citep{Johnson2023}, but poses significant challenges \citep{ngehtchallenge}. The imaging needs to deal with the relative sparsity of the array, various calibration issues at mm-wavelengths, super-resolution, polarimetry and time dynamics simultaneously. Several algorithms and softwares have been proposed that could deal specifically with time-dynamic reconstructions at the event horizon scales \citep{Bouman2017, Johnson2017, Arras2022, Mueller2022c, Mus2023} or static polarimetry \citep{Chael2016, Broderick2020, Broderick2020b, Vidal2021, Pesce2021, Mueller2022c}. Recently the potential of these approaches was demonstrated by \citet{ngehtchallenge}. However, the combined capability, i.e. the reconstruction of polarimetric movies within one framework, remains a unique imaging capability for only a few algorithms, such as DoG-HiT \citep{Mueller2022c}.

In this manuscript we build on the success of the recently proposed MOEA/D algorithm \citepalias{Mueller2023} in presenting a fast, self-calibration independent ({as long as self-calibration agnostic data functionals are used}) and unsupervised alternative to the already established imaging methods {in total intensity} and extend it both to polarimetric and time dynamic reconstructions in the same framework. In this spirit, this work contributes towards an automatic, unbiased reconstruction of polarimetric movies, particularly for the needs of the EHT and its planned successors.

\section{VLBI measurements and polarimetry}\label{sec:vlbibases}

The correlated signal of an antenna pair (the visibility $\mathcal{V}(u,v)$) is described by the van-Cittert-Zernike theorem \citep{Thompson2017}:
\begin{align}
    \mathcal{V} (u, v) \approx \int \int I(l, m) e^{-2 \pi i (l u + m v)} dl dm \,,  \label{eq: vis}
\end{align}
i.e. the true sky brightness distribution $I(l,m)$ and the visibilies form approximately a Fourier pair. The harmonic coordinates $(u,v)$ are determined by the baseline separating each antenna pair relative to the sky plane. Not all Fourier frequencies are measured, i.e. the $u,v$-coverage is sparse. Hence, imaging (i.e. recovering the sky brightness distribution from the observed visibilities) is an ill-posed inverse problem, since the codomain is only sparsely sampled. Furthermore, the visibilities are corrupted by calibration issues and thermal noise. We have to correct for these effects during imaging as well. The measured visibilities at a time $t$ observed along one pair of antennas $i,j$ are multiplied by station-based gain factors at each instant $t$, $g_{i,t}$ to correct direction-independent calibration effects, mathematically expressed as

\begin{align}
    V(i,j,t) = g_{i,t} g_{j,t}^{*} \mathcal{V}(i,j,t) + N_{i,j,t}\,,
\end{align}

with a baseline-dependent random noise term $N_{i,j,t}$. Hereafter, we consider $V,\mathcal{V},I,g$ and $N$ to be time-dependent, unless the contrary is said. For simplifying the notation, we omit the $t$.

The polarized fraction of the light recorded at every antenna is splitted by polarization filters (either orthogonal linear, or righthanded/lefthanded circular). These four channels per baseline (two per antenna) are combined in the four Stokes parameters: $I, Q, U, V$. $I$ describes non-negative total intensity, $Q, U$ linear polarization and $V$ the circular polarization. The Stokes parameters satisfy the following inequality:
\begin{align}
 I^2 \geq Q^2+U^2+V^2, \label{eq: stokes_inequality}
\end{align}
i.e. the total intensity of polarized and unpolarized emission is always greater than the polarized intensity. Similar as for Stokes I imaging we identify Stokes visibilities for the various Stokes parameters \citep{Thompson2017}:
\begin{align}
    &\mathcal{V}_I =  \mathcal{F}I, \\
    &\mathcal{V}_Q =  \mathcal{F}Q, \\
    &\mathcal{V}_U =  \mathcal{F}U, \\
    &\mathcal{V}_V =  \mathcal{F}V,
\end{align}
{where $\mathcal{F}$ denotes the Fourier transform.} We treat the Stokes visibilities as the observable for the remainder of this manuscript.

{Preparing the discussion of leakage corruptions, and the corresponding calibration within MOEA/D, we also discuss the formulation by visibility matrices here. However, note that we will fit the Stokes visibilities directly with MOEA/D, and not the coeherency matrices. }For circular feeds, we arrange the polarized visibilities for an antenna pair $i,j$ in the visibility matrix \citep{Hamaker1996}:
\begin{align}
\mathbf{\mathcal{V}}_{ij} = \left( \begin{array}{cc}
\mathcal{V}_I^{ij}+\mathcal{V}_V^{ij} & \mathcal{V}_Q^{ij}+i \mathcal{V}_V^{ij} \\
\mathcal{V}_Q^{ij}+i \mathcal{V}_V^{ij} & \mathcal{V}_I^{ij}-\mathcal{V}_V^{ij}
\end{array} \right)
\end{align}

For polarimetry, we have to deal with station-based gains and thermal noise in all bands. Additionally we have to deal with leakages and feed rotations in polarimetry. Gains are easiest represented by the Jones matrix \citep{Thompson2017} for antenna $i$:
\begin{align}
\mathbf{J}^i_{gain} = \left(\begin{array}{cc}
g_r & 0 \\
0 & g_l
\end{array}\right),
\end{align}
and feed rotations by the matrix $\theta$ \citep{Thompson2017}:
\begin{align}
\mathbf{J}^i_{rotation} = \left(\begin{array}{cc}
\exp(i \theta) & 0 \\
0 & \exp(-i \theta)
\end{array}\right).
\end{align}
The leakage between the perpendicular polarization filters introduces cross-terms \citep{Thompson2017}:
\begin{align}
\mathbf{J}^i_{leakage} = \left(\begin{array}{cc}
1 & d_r \\
d_l & 1
\end{array}\right).
\end{align}
Let us denote the complete corruption by $\mathbf{J}^i = \mathbf{J}^i_{gain} \mathbf{J}^i_{leakage} \mathbf{J}^i_{rotation} $. Then the observed visibility matrix $\mathbf{V}_{i,j}$ is:
\begin{align}
 \mathbf{V}_{i,j} = \mathbf{J}^i \mathcal{V}_{i,j} \left(\mathbf{J}^j \right)^\dag+\mathbf{N}_{i,j},
\end{align}
with thermal noise $\mathbf{N}_{i,j}$.

\section{MOEA/D}
MOEA/D is a multiobjective optimization technique that was originally proposed in \citet{Zhang2008, Li2009} for multiobjective optimization problems in a general setting. The framework was adapted for VLBI imaging in \citetalias{Mueller2023}. In \citetalias{Mueller2023} we connected the various objectives that are balanced against each other in a RML formulation of the imaging problem, i.e. the various data terms and penalizations, to the single objectives of a multiobjective problem formulation. In this way, MOEA/D explores all possible regularization parameter combinations simultaneously, and a lengthy parameter survey could be omitted, although the number of pixels is a crucial factor in the complexity of the problem. A new technique not as sensitive to the number of pixels based on {particle swarm optimization} is being developed~\cite{Mus2023b}. Moreover, we introduced a more global search technique by the genetic algorithm in \citetalias{Mueller2023}. In this section, we recall some of the basic concepts of MOEA/D. For more details we refer to \citetalias{Mueller2023}.

For MOEA/D, the imaging problem is reformulated as a multiobjective optimization problem, i.e. as the problem \citepalias{Pardalos2017, Mueller2023}:
\begin{equation}
  \label{prob:mop}
  \tag{$\text{MOP}$}
    \begin{aligned}
      & \underset{x\in D}{\text{min}}
      & & F\left(x\right) := \left(f_1\left(x\right),\ldots,f_n\left(x\right)\right),\\
      & \text{subject to}
      & & x\in D\subset\RR^m,\\
    \end{aligned}
\end{equation}
where $f_i:D\longrightarrow\RR,\ i=1\ldots,m$ are the single objective functionals, $D$ is the decision space, and $F:D\longrightarrow\RR^n$ the vector-valued multiobjective optimization functional. For the minimization of $F$ we look for best compromise solutions. A guess solution $x^* \in D$ is called Pareto optimal or nondominated, if the further optimization in one objective automatically worsens the scoring in another objective functional, i.e. if no point $x\in D$ exists, such that $f_i\left(x\right)\leq f_i\left(x^*\right),\ \forall i=1,\ldots,m$ and $f_j\left(x\right)<f_j\left(x^*\right)$ for at least one $j=1,\ldots,m$ \citepalias{Pardalos2017, Mueller2023}. The set of all Pareto optimal solutions is called the Pareto front \citep{Pardalos2017}.

In \citetalias{Mueller2023} we chose the following modelization for Stokes I imaging:
\begin{align}
    & f_1:=\alpha S_{vis}+\beta S_{amp}+\gamma S_{clp}+\delta S_{cla}+\zeta R_{l1},\\
    & f_2:=\alpha S_{vis}+\beta S_{amp}+\gamma S_{clp}+\delta S_{cla}+\theta R_{tv},\\
    & f_3:=\alpha S_{vis}+\beta S_{amp}+\gamma S_{clp}+\delta S_{cla}+\tau R_{tsv},\\
    & f_4:=\alpha S_{vis}+\beta S_{amp}+\gamma S_{clp}+\delta S_{cla}+ \eta R_{l2},\\
    & f_5:=\alpha S_{vis}+\beta S_{amp}+\gamma S_{clp}+\delta S_{cla}+\epsilon R_{flux},\\
    & f_6:=\alpha S_{vis}+\beta S_{amp}+\gamma S_{clp}+\delta S_{cla}+\kappa R_{entr},\\
    & f_7:=\alpha S_{vis}+\beta S_{amp}+\gamma S_{clp}+\delta S_{cla},
\end{align}
where $S_{vis}, S_{amp}, S_{clp}, S_{cla}$ are the reduced $\chi^2$-metrics with respect to the visibilities, amplitudes, closure phases and closure amplitudes respectively, and $R_{l1}, R_{l2}, R{tv}, R_{tsv}, R_{flux}, R_{entr}$ denote the $l^1$-, $l^2$-, total variation, total squared variation, compact flux and entropy penalty terms. For more details on these terms we refer to the more detailed discussion in \citetalias{Mueller2023}.

{The output of the MOEA/D is a sample of potential image features covering the decision space among the axes of all data terms and regularizers rather than a single image. Hence, the problem is high-dimensional limiting the number of pixels to achieve a sufficient numerical performance. Furthermore, the genetic algorithm, while exploring the parameter space globally, does not take into account gradient or hessian information, thus converges rather slow. These points are tackled in a companion effort \citep{Mus2023b} that replaces the genetic evolution by a hybrid approach and waives the limitation of a small number of pixels.}

Some of these objectives have discrepancies. For instance sparsity in the pixel basis (promoted by $R_{l1}$) and smoothness (promoted by $R_{tsv}$) are contradicting assumptions. In essence, this means that a single image that minimizes all objective functionals $f_i$ typically does not exist \citep{Pardalos2017}. For RML methods we achieve a potential, regularized reconstruction by balancing these terms with the proper, but initially unknown, weighting parameters $\alpha, \beta, ...$ by a weighted sum minimization. For MOEA/D all the different weighting combinations are explored simultaneously and are approximated by the Pareto front \citep{Zhang2008}. Due to the aforementioned discrepancies between data terms and penalty terms and in between different penalty terms, the Pareto front divides into several clusters of solutions \citepalias{Mueller2023}. The number of disconnected clusters is larger, the weaker the image features are constrained, i.e. for enhanced sparsity of the array and for self-calibration independent closure-only imaging \citepalias{Mueller2023}. The recovered image morphologies within one cluster vary only marginally. However, the recovered image features among several disconnected clusters vary significantly. We identify the independent clusters in the Pareto front by a friends-on-friends algorithm \citepalias[for more details we refer to][]{Mueller2023}.

While all the nondominated solutions are explored simultaneously within the Pareto front, we have to deal with the problem to find the cluster of solutions that is objectively best. In \citetalias{Mueller2023} we argued that this could be done without the need for lengthy parameter surveys by a more automatized choice in a data-driven way. We proposed to use the solution that has the highest number of close neighbors (i.e. the dominating cluster of solutions), or the cluster that is minimizing the (euclidean) distance to the ideal point (mimicking a least-square principle). In the application to synthetic data and to real data, particularly the choice with the largest number of close neighbors performed best \citepalias{Mueller2023}.

Generally, it is challenging to compute the Pareto front analytically. Hence, the Pareto front is typically approximated by a sample of characteristic members \citep{Pardalos2017}. Several strategies exist to approximate the Pareto front. Most of these algorithms decompose the multiobjective optimization problem into a sample of single-objective problems either by the Tchebycheff Approach, the Boundary Intersection Method, or the weighted sum approach \citep{Tsurkov2001, Zhang2008, Xin2019, Sharma2022}. We applied the weighted sum approach in \citetalias{Mueller2023} due to its philosophical similarity to parameter surveys that are common in VLBI for RML methods \citep[e.g. compare][]{eht2019d, Kim2020, Janssen2021, eht2022c, Zhao2022, Fuentes2022}. In this way, the Pareto front approximates the set of images using parameter surveys. We introduced normalized weight arrays $\lambda^1 = \{\lambda_0^1, \lambda_1^1, ..., \lambda_m^1\}, \lambda^2=\{\lambda_0^2, \lambda_1^2, ..., \lambda_m^2\}, ...$ that are related to the objective functionals $f_1, f_2, ..., f_m$ in \citetalias{Mueller2023} to rewrite as a weighted sum the multiobjective formulation of Prob.~\eqref{prob:mop} into a sequence of single-objective optimization problems~\citepalias{Mueller2023}:
\begin{align}
x^j \in \mathrm{argmin}_{x} \sum_{i=1}^m \lambda_i^j f_i(x)\,.
\end{align}
{Note that there are two different weights. The weights $\alpha, \beta, ...$ (which remain fixed during the running time of MOEA/D) and the weights $\lambda^i$ (which we search over for Pareto optimal solutions). The weights $\alpha, \beta, ...$ are used to renormalize the regularization terms and data terms to a similar order of magnitude to help the convergence procedure. However, they do not affect the estimation of the Pareto front as long as the grid the weights $\lambda^i$ is large and dense enough.\citepalias[for more details see Appendix A in][]{Mueller2023}.}

One particularly successful strategy to approximate the Pareto front by decomposition is the multiobjective evolutionary algorithm MOEA/D proposed in \citet{Zhang2008, Li2009}. MOEA/D computes the Pareto front by a genetic algorithm. We start with a random initial population. The subsequent population is obtained by evolutionary operations, i.e. by random mutation and genetic mixing. MOEA/D fits specifically well into the framework of VLBI imaging, since only parent solutions with similar weight combinations $\{\lambda_0^j, \lambda_1^j, ..., \lambda_m^j\}$ are genetically mixed, i.e. the algorithm keeps the diversity within the population \citepalias{Zhang2008, Mueller2023}.

MOEA/D for VLBI combines several advantages. It is faster than complete parameter surveys for RML methods or exact Bayesian sampling schemes, provides an alternative claim of the image structure in an automatized (unsupervised) way, and explores the image features with a more global search technique, i.e. by a randomized evolution in a multiobjective framework \citepalias{Mueller2023}. In conclusion, the {dependence on the selection of regularization terms} inherent to all {RML} imaging procedures is addressed effectively by a multiobjective, more global exploration of the possible solution space.

\section{Modelization of the Problem}

\subsection{Polarimetry}
For polarimetry, we introduce the $\chi^2$-fit to the linear polarized visibilities $\mathcal{V}_Q, \mathcal{V}_U$ as an additional data term: $S_{pvis}$. {Although it may be a small extension to also include circular polarization in the set of objectives and recover circular polarization among the linear polarization, we omit to this extension. Recovering circular polarization requires a range of additional calibration steps, particularly of the R/L gain ratio that are typically not performed in self-calibration, but across multiple sources \citep{Homan1999, Homan2006}.}

Moreover, we use specifically designed polarimetric regularization terms inspired by the selection in \citet{Chael2016} and subsequently used by the EHT collaboration \citep[e.g.][]{eht2021a}. In particular, we use the functionals $R_{ms}, R_{hw}, R_{ptv}$ that will be presented in the following paragraphs. 

First, we use the conventional KL polarimetric entropy~\citep{Ponsonby1973,Narayan1986,Holdaway1988,Chael2016} of an image $I$ with respect to a prior image $M_i$ (chosen to be a Gaussian with roughly the size of the compact flux region):
\begin{align} \nonumber
    R_{hw} := \displaystyle{\sum^{N^2}_{i}} I_i \left( \ln(\dfrac{I_i}{M_i}) + \dfrac{1+m_i}{2} \ln\left(\dfrac{1+m_i}{2}\right) \right.\\
    \left. + \dfrac{1-m_i}{2} \ln\left(\dfrac{1-m_i}{2}\right)\right). \label{eq:shw}
\end{align} 
Here $m_i$ denotes the fraction of linear polarization in pixel $i$. $R_{hw}$ favors images of $N$ pixels with fractional polarization, $\Vec{m}$ less than one.

The second regularizer $R_{ms}$ {extends the concept of the entropy to the polarized image:}
\begin{equation}\label{eq:msimple}
    R_{ms} = \displaystyle{\sum^{N}_{i}}\lvert I_i\rvert \ln \lvert m_i \rvert.
\end{equation}
{We note that $R_{ms}$ diverges to negative infinity if the polarization fraction approaches zero. To avoid this, we atcually use $\mathrm{max}(R_{ms}, -z)$ as regularization term. Here $z$ is a real number that has been found as the smallest scroring $R_{ms}(x)$ in the Pareto front that has not diverged.}
Finally, we utilize the polarimetric counterpart of the total variation regularizer \citep{Chael2016}:
\begin{equation}\label{eq:tv}
    R_{ptv} = \sum_{i}\sum_{j}\sqrt{\vert P_{i+1, j} - P_{i,j} \rvert^2 + \left| P_{i, j+1} - P_{i,j} \right|^2},
\end{equation}
where we used the complex realization of the linear polarized emission $P = Q+iU$ to note down the total variation for $Q$ and $U$ in a compact form.

In conclusion, the resulting multiobjective problem consists of the single functionals: 
\begin{align}
    & f_1:=\alpha S_{pvis}+\beta R_{ms}, \label{eq: pol1}\\
    & f_2:=\alpha S_{pvis}+\gamma R_{hw},\label{eq: pol2}\\
    & f_3:=\alpha S_{pvis}+\delta R_{ptv},\label{eq: pol3}\\
    & f_4:=\alpha S_{pvis}. \label{eq: pol4}
\end{align}

\subsection{Dynamics}
The philosophy behind time dynamic imaging is as follows: We cut the full observation into $r$ keyframes where, for a given time window, all frames satisfy that there is at least one visibility observed. The visibilities of one frame are:
\begin{equation}\label{eq:keyframes}
\mathcal{V}^l = \left\{\mathcal{V} ~\mathrm{for}~~ t \in [t_l - \Delta t_l/2, t_l + \Delta t_l/2] \right\}, ~~ l \in [1,r],
\end{equation}
where $t^l$ is the observing time, and $\Delta t_l$ is a frame-dependent scalar that determines its duration. For each $l$, the associated data set $\mathcal{V}^l$ produces an ``image keyframe''. The model will have a total of $r \times N^2$ parameters (i.e., $r$ images of $N^2$ pixels each). Now we naturally extend the data terms and penalty terms to a time-series, e.g. by:
\begin{align}
S_{vis}(\mathcal{V}, p) = \frac{1}{r} \sum_{l=1}^r S_{vis}(\mathcal{V}^l, p^l),
\end{align}
where $p$ is a time-series of images (i.e. a movie). We proceed analogously for all the other data terms and regularization terms.

In~\citetalias{Mueller2023}, we constructed Prob.~\eqref{prob:mop} using seven functionales including $l^2$-norm ($R_{l2}$), $l^1$-norm ($R_{l1}$), total variation (TV) and total squared variation (TSV)($R_{tv}, R_{tsv}$), flux modeling ($R_{flux}$), the entropy ($R_{entr}$) and standard visibility and closures data terms ($S_{vis},S_{amp},S_{clp},S_{cla}$) (detailed descriptions of each regularizer can be found in the referenced paper.). 

To extend this problem to include dynamics we consider a new objective $f_{\mathrm{ngMEM}}$, the ngMEM regularization term defined by \citep{Mus2023} acting on a time-series of images $p \in \RR^{r \times N^2}_{+}$: 
\begin{equation} \nonumber
\mu_{\mathrm{ngMEM}}R_{\mathrm{ngMEM}} = \sum_{n, j \neq k}{T_n^{jk}},
\end{equation}
where  $j,k \in \{1, 2, ..., r\}$ denote the frame in the time-series, $\mu_{\mathrm{ngMEM}}$ is a hyper-parameter and
\begin{equation}
\label{eq:image_memory}
    T_n^{jk} := 
    e^{-\frac{|t^j - t^k|^2}{2\pi^2}}
    \left(|p_n^j - p_n^k |+C\right)\log{\left( |p_n^j - p_n^k| + C\right)}.
\end{equation}
Here, $t^j$ denotes the times of the associated keyframe $j$. This functional seeks the most constant dynamic reconstruction (or movie hereafter) which still fits into the data \citep{Mus2023}.

The ngMEM is using the Shannon entropy~\cite{Shannon1949} without imposing any a priori model to ensure a uniform distribution in both brightness and time for a time-series of such frames. Therefore it is a conservative approach, since any change in contrast will contribute to the increment of the entropy.

The two parameters $\mu_{\mathrm{ngMEM}}$\footnote{To keep consistency between the notation of the bibliography and this paper, we have denoted by $\mu_{\mathrm{ngMEM}}$ the weight associated to the ngMEM and $\mu$ the normalized weight considered for the associated MOP.} and $\pi$ are used as regularizers and must be explored. $\pi$, the ``time memory'', has a minor effect, while $\mu_{\mathrm{ngMEM}}$, the ``time weight'', must be carefully selected. Therefore, $\mu_{\mathrm{ngMEM}}$ should be investigated using MOEA/D. {Finally, $C$ is just a floor that should be small to avoid extremely large values or not-a-number errors for the logarithm.}

We can solve this problem of dynamic imaging (i.e., find the set of image keyframes that optimally fit the data) by using the following formalism. 
\begin{align}
    & f_1:=\alpha S_{vis}+\beta S_{amp}+\gamma S_{clp}+\delta S_{cla}+\zeta R_{l1}, \label{eq: dyn1}\\
    & f_2:=\alpha S_{vis}+\beta S_{amp}+\gamma S_{clp}+\delta S_{cla}+\theta R_{tv},\label{eq: dyn2}\\
    & f_3:=\alpha S_{vis}+\beta S_{amp}+\gamma S_{clp}+\delta S_{cla}+\tau R_{tsv},\label{eq: dyn3}\\
    & f_4:=\alpha S_{vis}+\beta S_{amp}+\gamma S_{clp}+\delta S_{cla}+ \eta R_{l2},\label{eq: dyn4}\\
    & f_5:=\alpha S_{vis}+\beta S_{amp}+\gamma S_{clp}+\delta S_{cla}+\epsilon R_{flux},\label{eq: dyn5}\\
    & f_{6}:=\alpha S_{vis}+\beta S_{amp}+\gamma S_{clp}+\delta S_{cla}+ \kappa R_{\mathrm{entr}}, \label{eq: dyn6}\\ 
    & f_7:=\alpha S_{vis}+\beta S_{amp}+\gamma S_{clp}+\delta S_{cla} + \mu R_{\mathrm{ngMEM}}, \label{eq: dyn7} \\
    & f_8:=\alpha S_{vis}+\beta S_{amp}+\gamma S_{clp}+\delta S_{cla}.\label{eq: dyn8}
\end{align}

\subsection{Dynamic polarimetry}

Ultimately, the fusion of polarimetry regularizers with ngMEM offers a potent approach for the retrieval of polarimetric evolution. Given the inherent intricacies of this challenge, only few algorithms possess the capability to address it. In particular, the only RML-like algorithm that is currently able to do so is DoG-HiT~\citet{Mueller2022c}. The Bayesian algorithm \texttt{resolve}~\cite{Arras2019,Arras2021} has the potential to also recover dynamic polarimetry, although no publication up-to date has been done in that direction.

The modelization of the problem within the framework of MOEA/D is notably straightforward. In principle this could be achieved in the most consistent way by combining Eq. \eqref{eq: pol1}-\eqref{eq: pol4} with Eq. \eqref{eq: dyn1}-Eq. \eqref{eq: dyn7}, i.e. by adding the polarimetric objectives to every single scan and solving for total intensity and polarization at the same time. However, the problem becomes rapidly high-dimensional and complex to solve, particularly when exploring the {range of possible solutions} the Pareto front. {In particular, the complexity of the MOEA strategies ranges between $O\left(n_on_{p}^{2}\right)$ and $O\left(n_on_{p}^{3}\right)$~\citep{curry2014}, being $O$ the asymptotic upper bound. The population size is the number of weight combinations drawn from an equidistant grid with overall sum 1, i.e. $n_p = \mathrm{binom}(n_o+n_g-1,n_g)$ where $n_g$ is the grid size. Current work is ongoing to achieve a better scaling by a grid-free evolution of the weights \citep{Mus2023b}.}

{Moreover, since gains and d-terms are not part of the forward model, an application in practice is naturally limited to aa hierarchical approach: solve for the d-terms and gains from the static image (with subsequent refinement with the movie in total intensity first) before proceeding to a full polarimetric movie reconstruction.}

Therefore, we split the strategy into several steps. First, we solve for the dynamics in total intensity as described above (step 1). Second we calculate a static polarimetric image as described above (step 2). Third, we cut the observation into keyframes and solve the polarimetric imaging at every keyframe independently with MOEA/D, initializing the population with the final population of the static polarization step (step 2) and assuming the Stokes I image from the time-dynamic exploration (step 1). This strategy resembles the strategy that was also applied in \citet{Mueller2022c} for DoG-HiT, one of the few algorithm that notably has the capability for dynamic polarimetry already.

\subsection{Pareto Fronts}
In case of polarimetry, the Pareto front is a four-dimensional hypersurface (four objective terms). As for total intensity, the Pareto front divides into several disjunct clusters due to conflicting assumptions of several regularization terms. We examine the Pareto front in the same way as described in \citetalias{Mueller2023}, i.e. we identify the clusters of solutions in the scoring by a friends-on-friends algorithm. Afterwards, we evaluate the representative solutions of every cluster by its number of close neighbors and by the distance to the ideal, for more details see \citetalias{Mueller2023}. For special clusters, i.e. for the clusters that present overfitted data, the polarization fraction {accumulates at the edges of the interval [0,1]}. In this case, the regularizer $R_{ms}$ approaches negative infinity. We explicitly exclude these clusters for the finding of the objectively best solution. 

In the dynamic case, the dimensionality of the Pareto front is 8. The resulting ``Pareto movies'' are defined by the action of the different objectives on the recovered movies defined by the frames obtained.

Finally, the Pareto front obtained solving the dynamic polarimetric problem is composed by the two disjoint Pareto fronts defined above.

\section{Verification on synthetic data in EHT+ngEHT array}\label{sec:synthetic_data}

\subsection{Array Configurations and Problem Framework}
\begin{figure}
    \centering
    \includegraphics[width=0.5\textwidth]{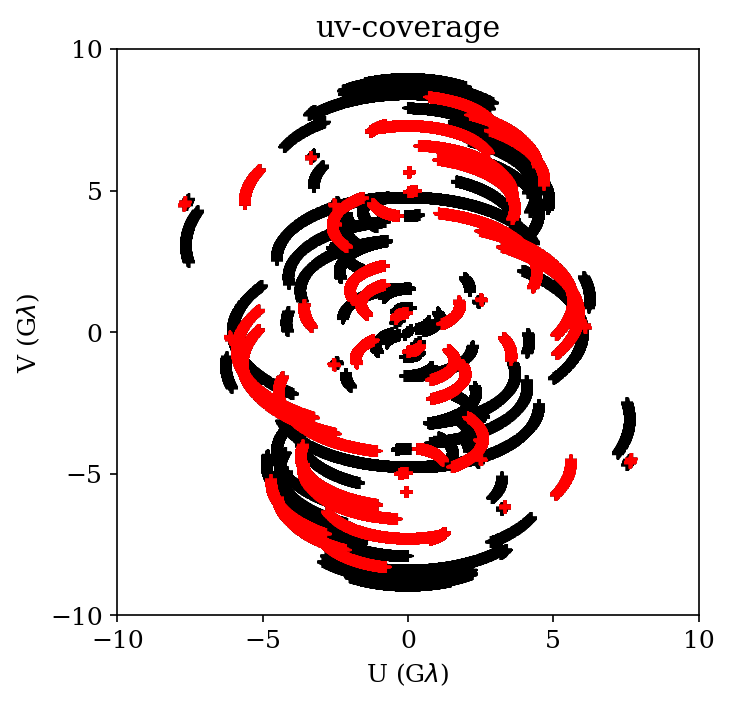}
    \caption{uv-coverage for \sgra in 2017 EHT (red dots) and ngEHT at 230\,GHz (black crosses).}
    \label{fig:uv_cov}
\end{figure}

The self-consistent reconstruction of polarimetric dynamics at the event horizon scales is one of the major goals of the EHT and its planned successors \citep{Johnson2023, ngehtchallenge}. Hence, in consistency with our analysis presented in \citet{Mueller2023}, we study two array configurations for the remainder of this manuscript. On one hand, we study the configuration of the EHT observations of SgrA* in 2017 at $230\,\mathrm{GHz}$ \citep{eht2022a}. On the other hand, we study a possible EHT + ngEHT configuration that was also used within the ngEHT Analysis challenges \citep{ngehtchallenge}, including ten additional stations at $230\,\mathrm{GHz}$, a quadrupled bandwidth and an enhanced frequency coverage. The corresponding uv-coverages are shown in Fig. \ref{fig:uv_cov}.

For consistency with~\citetalias{Mueller2023}, we have solved the different MOPs using 256 pixels.

\subsection{Polarimetry}
\begin{figure*}
    \centering
    \includegraphics[width=\textwidth]{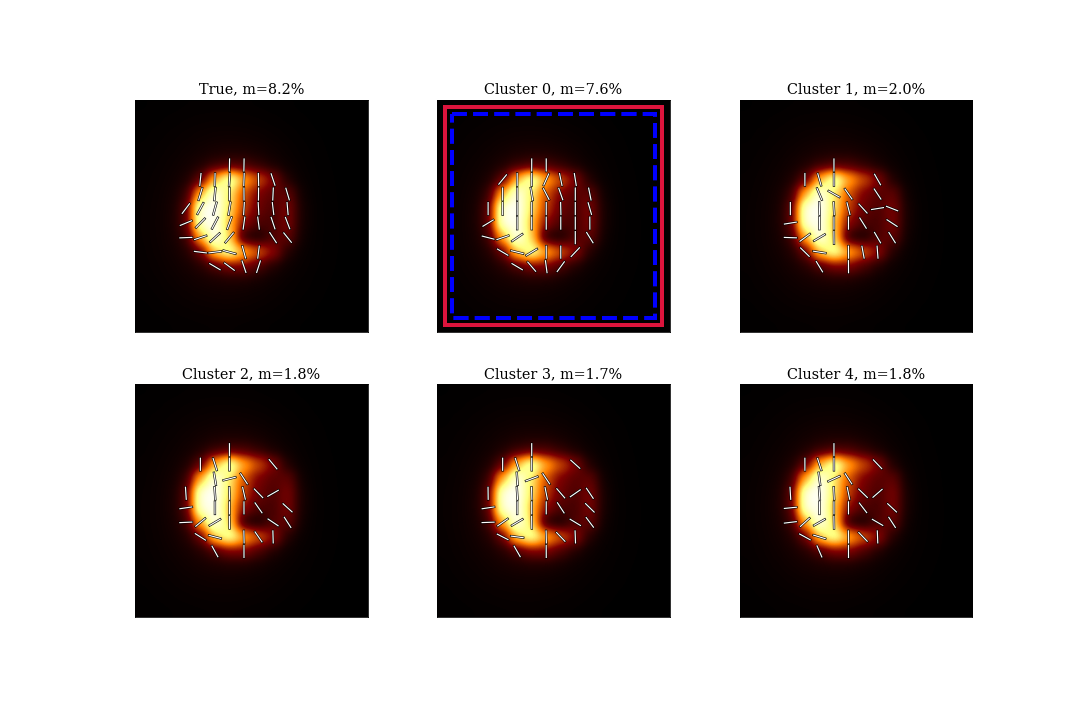}
    \caption{Clustered images for simulated \sgra polarimetric model using the EHT 2017 array at 230 GHz and. The preferred cluster by the largest number of representants is indicated by a red box, the one preferred by the closest distance to the ideal point by a blue box. Vertical lines represent the EVPA}.
    \label{fig: pol_eht_sgra}
\end{figure*}

To demonstrate the capabilities of MOEA/D in the polarimetric domain, we use a synthetic data source out of the \textit{ehtim} software package \citep{Chael2016, Chael2018} that was previously used in \citet{Mueller2022c} for synthetic data tests as well. The ground truth image is plotted in the upper left panel of Fig. \ref{fig: pol_eht_sgra}. The model mimics the typical crescent-like black hole shadow of \sgra in total intensity with a highly linear polarized structure.

We show our reconstructions in the same way as we did in \citetalias{Mueller2023}. The non-dominated solutions in the final, evolved population of the Pareto front, form disconnected clusters of solutions. The variety of the recovered structure within one cluster is small, the variety in the recovered image structures between different clusters of reconstructions however is significant. We present in Fig. \ref{fig: pol_eht_sgra}, one representant of each cluster of solutions (the accumulation point) in comparison to the true image structure. {The existence of multiple clusters with a single particle starting point already hints that MOEA/D at least partially overcame local convexity and searched the solution space globally independent of the starting point.}

For the polarimetric reconstruction, we fixed the Stokes I reconstruction {since we suppose datasets need to be d-term calibrated (however note that if d-terms are lower than $~10\%$, imaging algorithms have stronger effects on the reconstruction than the d-terms~\citealt{Marti2008})}, and only solved for linear polarization with MOEA/D. {Moreover, this strategy mimics the strategy of splitting the RML parameter surveys into a total intensity and polarimetry part as was applied by the EHT.} We initialized the initial population with images with constant electric vector position angle (EVPA) at a constant polarization fraction of $1\%$ across the whole field of view. Rather than with the Stokes parameters $Q, U$, we equivalently model the linear polarization by the linear polarization fraction $m$ and the mixing angle $\chi$, i.e.:
\begin{align}
    Q + iU = m \cdot I \cdot e^{2\pi i \chi}.
\end{align}
This turns out to be beneficial since inequality \eqref{eq: stokes_inequality} is automatically satisfied by imposing $0 \leq m \leq 1$ {(assuming small Stokes $V$)} {during the genetic evolution. We recover some undesired solutions when the weighting favors $R_{ms}$ as this term promotes small polarization fraction approaching $m=0$, as shown in Fig.~\ref{fig: pol_eht_sgra}). However, these solutions are disfavored as they are at the edge of the Pareto front, and are not selected by neither of the selection heuristics}. Specifically we cannot create spurious polarimetric signals outside of the total intensity contours. MOEA/D evolves the population of solutions by genetic operations, i.e. by genetic mixing and random mutation. For latter step, it is beneficial for the numerical performance to rescale all guess arrays (i.e. $m$ and $\chi$) such that the resulting values are of order of 1, we refer to \citetalias{Mueller2023} for more details on the numerical performance.

Out of the clusters of solutions presented in Fig. \ref{fig: pol_eht_sgra} one (cluster 0) recovers the polarization fraction well. The EVPA pattern, particularly the small-scale perturbation towards the south-east of the ring, is recovered well. Cluster 0 is both strongly preferred by the absolute $\chi^2$ (i.e. it is the cluster that fits the data best) as also by the accumulation point criterion that we proposed in \citetalias{Mueller2023} (i.e. it is the cluster with the largest number of neighbors, hence the most 'common' one). 

\begin{figure*}
    \centering
    \includegraphics[width=0.75\textwidth]{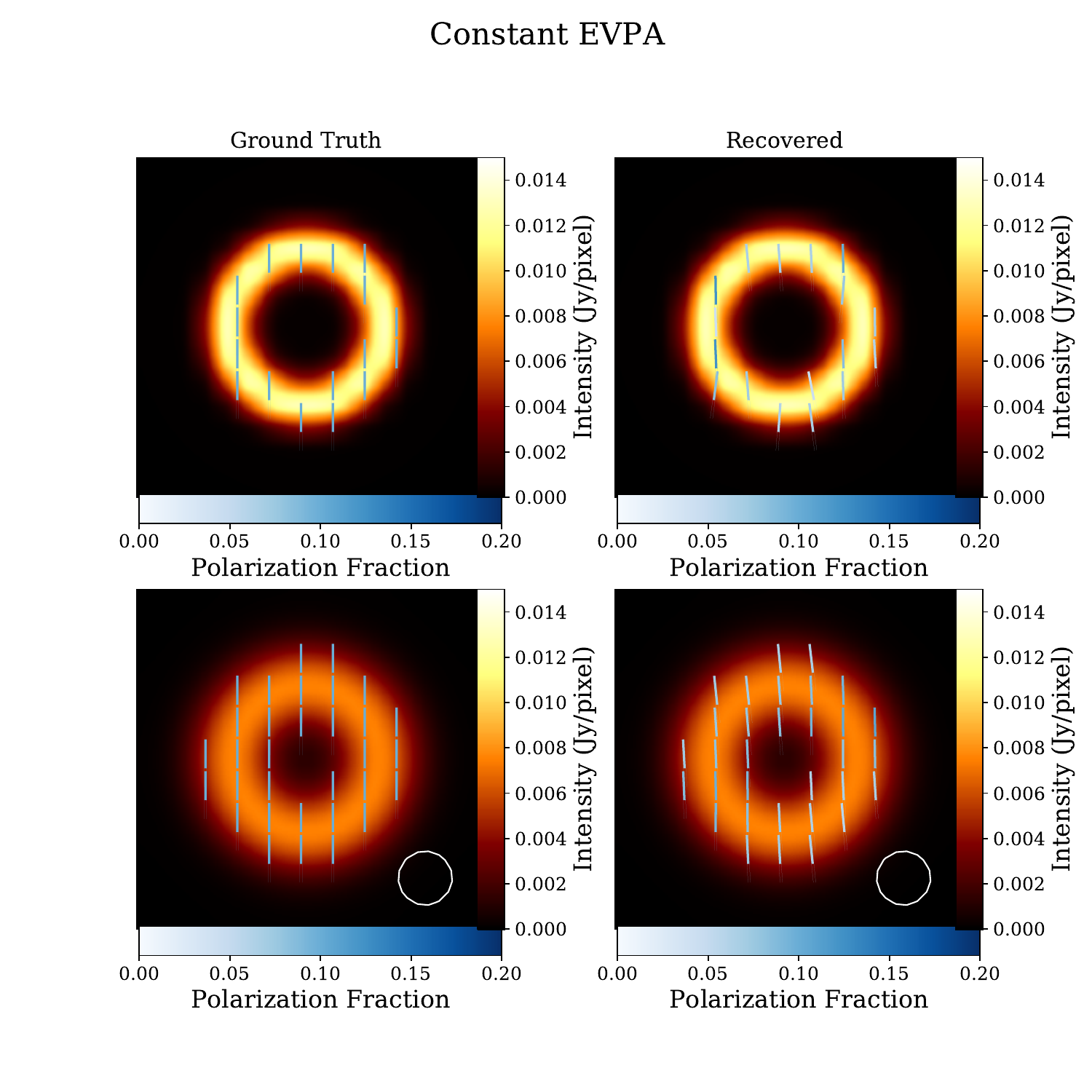}
    \caption{Simulated ring-like source with constant vertical EVPA using EHT 2017 array. Left panel: Ground truth. Right panel: preferred image (solution) recovered with MOEA/D image. {Bottom row: super-resolved structures. Bottom row: images convolved with 20\,$\mu$as beam shown in white.} Vertical lines represent the EVPA and their color, the amount of polarization fraction. The color map for the brightness of the source is different to the one of the EVPAs.}
    \label{fig: constant}
\end{figure*}

\begin{figure*}
    \centering
    \includegraphics[width=0.75\textwidth]{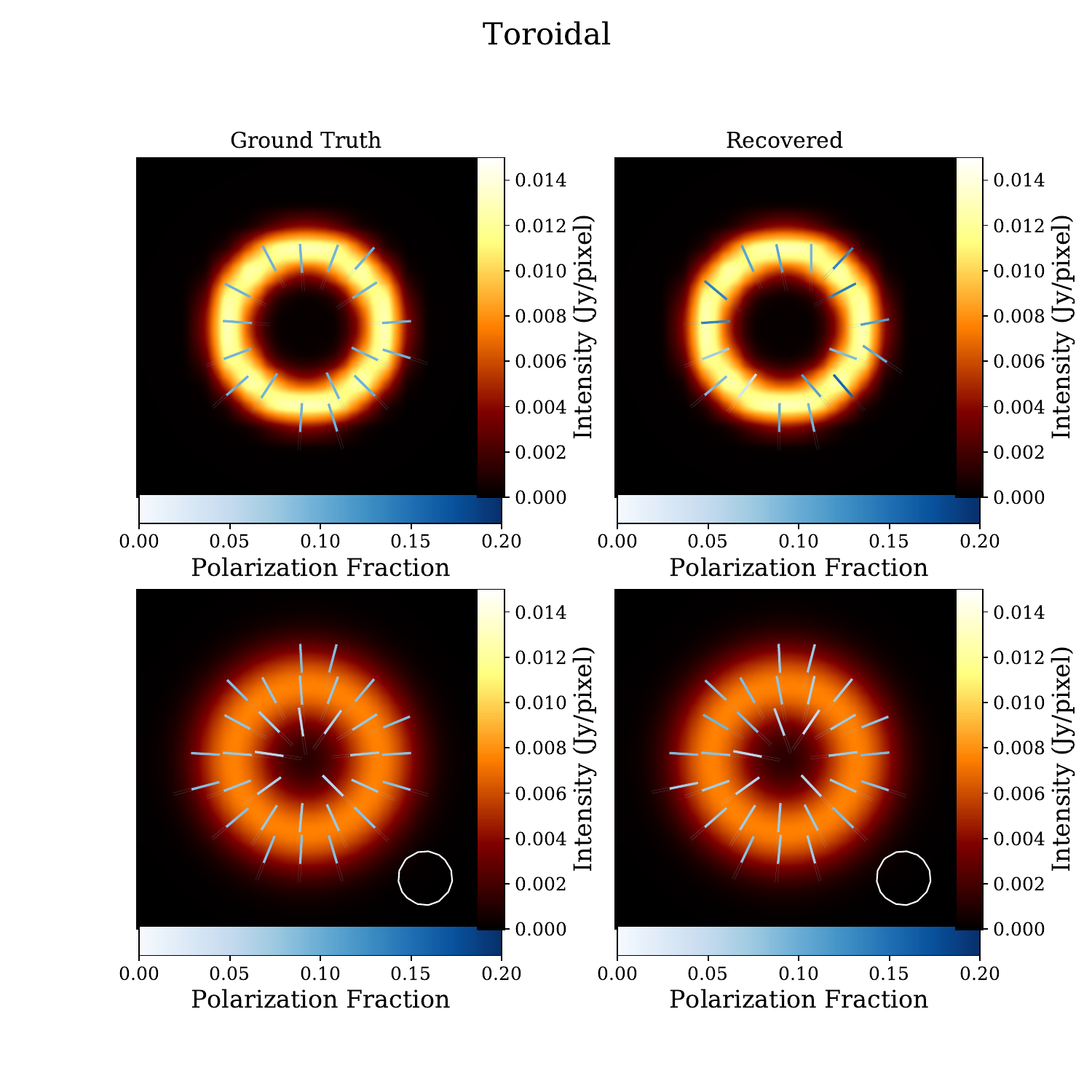}
    \caption{Same as Fig,~\ref{fig: constant} but the EVPA are following a toroidal magnetic field configuration.}
    \label{fig: toroidal}
\end{figure*}

\begin{figure*}
    \centering
    \includegraphics[width=0.75\textwidth]{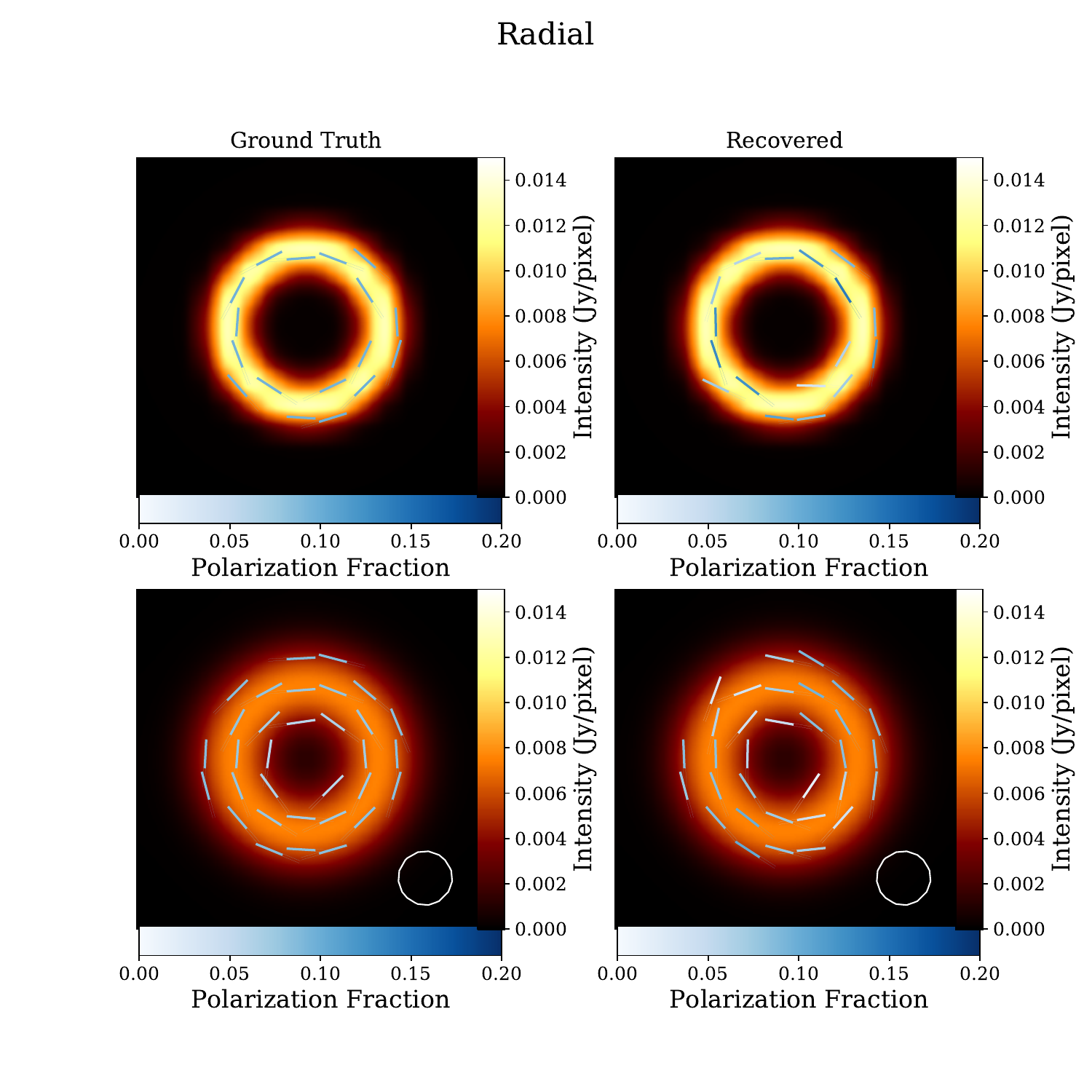}
    \caption{Same as Fig,~\ref{fig: constant} but the EVPA are following a radial magnetic field configuration.}
    \label{fig: radial}
\end{figure*}

\begin{figure*}
    \centering
    \includegraphics[width=0.75\textwidth]{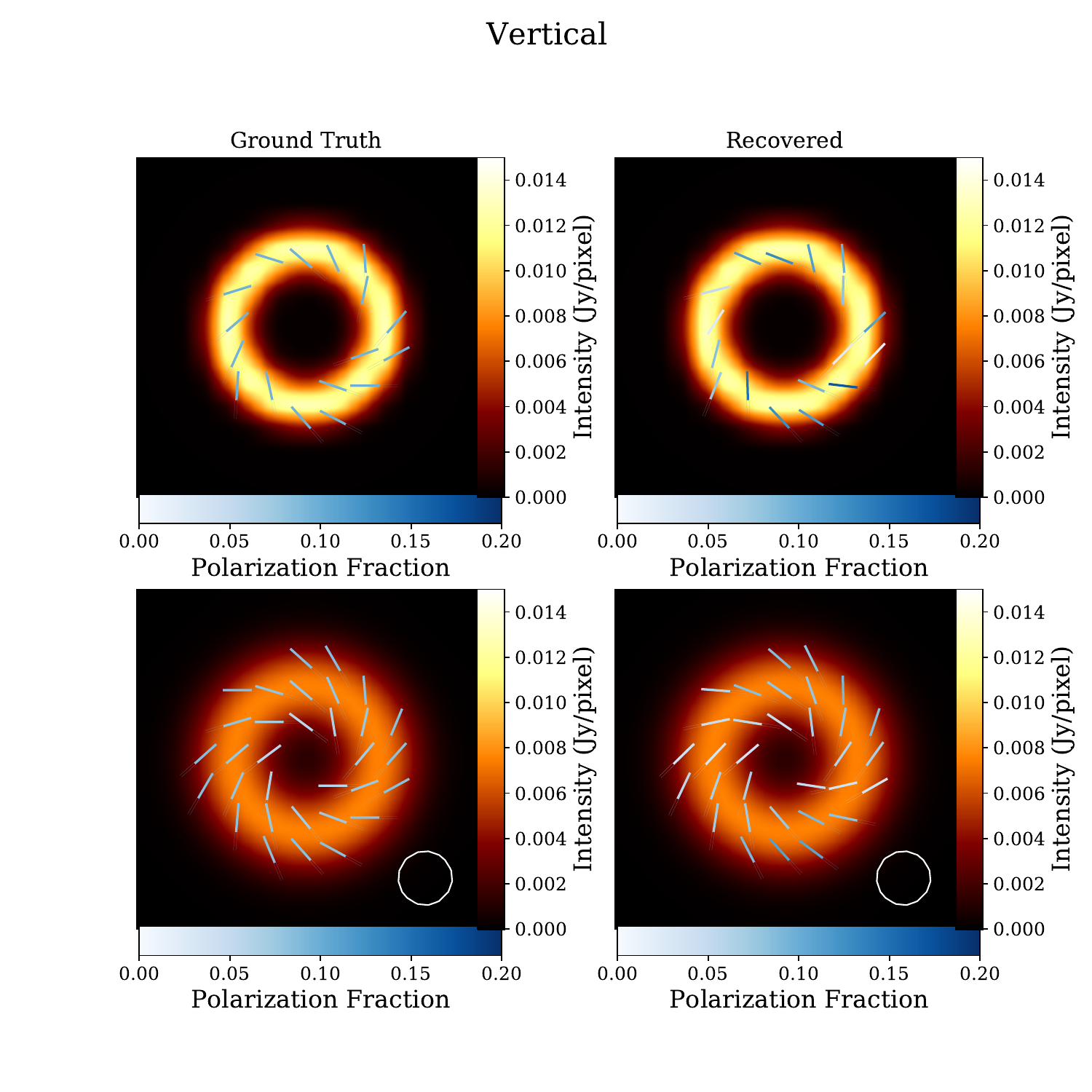}
    \caption{Same as Fig,~\ref{fig: constant} but the EVPA are following a vertical magnetic field configuration.}
    \label{fig: vertical}
\end{figure*}

To verify that MOEA/D reconstructions performs well for a wider range of EVPA structures on such a challenging data set as the EHT observations, we performed a number of additional tests with an artificial ring image and a constant EVPA pattern, a toroidal magnetic field, a radial magnetic field and a vertical magnetic field, see Fig. \ref{fig: constant}, Fig. \ref{fig: toroidal}, Fig. \ref{fig: radial}, and Fig. \ref{fig: vertical} {respectively}. {For each case,}  we show the respective best solutions selected by the accumulation point criterion {(right columns) and the real model (left columns). Bottom rows show model and solution convolved with a 20\,$\mu$as beam (represented as a white circle in the plot). The reconstruction of the overall pattern is successful in all four convolved cases, while ee observe how super-resolution scrambles the pattern}.  In conclusion, we can differentiate various magnetic field configurations by the MOEA/D reconstructions of EHT like data.

\begin{figure*}
    \centering
    \includegraphics[width=\textwidth]{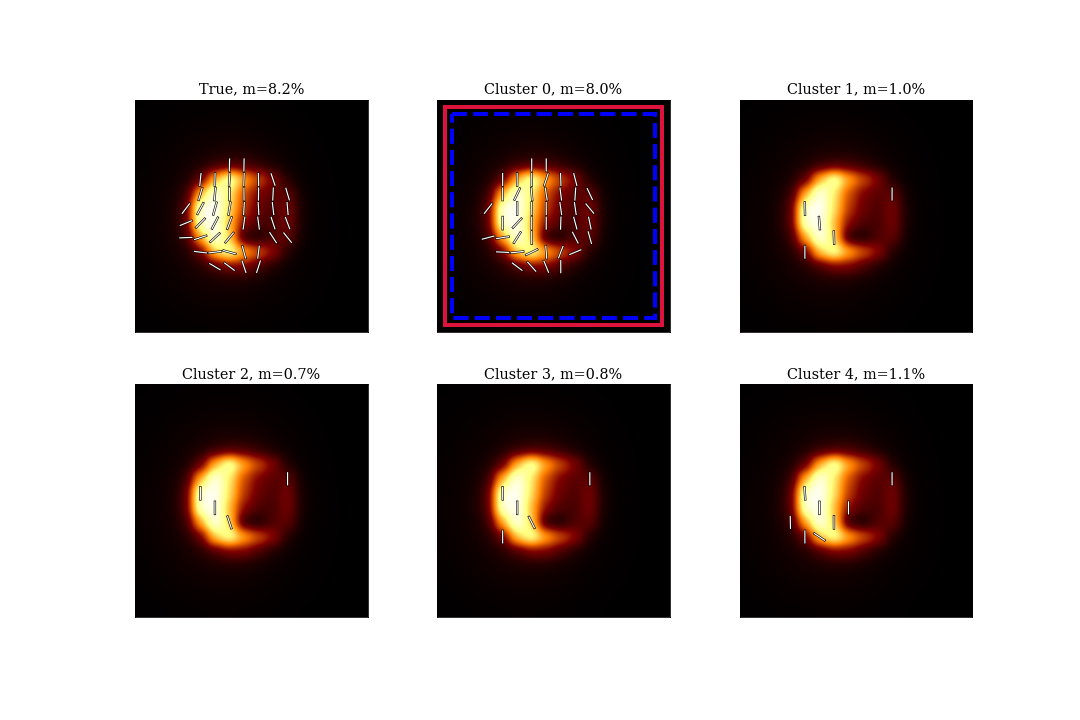}
    \caption{Cluster of images obtained with MOEA/D for a simulated polarimetric \sgra model using EHT+ngEHT array configuration at 230\,GHz.}
    \label{fig: pol_ngeht_sgra}
\end{figure*}

Now we study how the reconstruction improves with an improved array. We present the same GRMHD {simulated} reconstruction observed with a ngEHT configuration in Fig. \ref{fig: pol_ngeht_sgra}. Due to the larger number of visibilities, the observation is more constraining. The reconstruction within the preferred cluster(cluster 0) improves, particularly with respect to the total polarization fraction. Interestingly, the non-preferred clusters (at the edge of the parameter space) still exist, {resembling the weight combinations that overweight $R_{ms}$}, but are strongly discouraged by investigating the properties of the Pareto front. For more details regarding tools to investigate the Pareto front, e.g. by looking for disjoint clusters, accumulation points and the solution closest to the optimum, we refer to \citetalias{Mueller2023}.

Finally, we study the reconstruction performance in the presence of more realistic data corruptions as they may be expected in real data sets. For this we introduced gain errors (i.e. the need for self-calibration) and leakage errors (i.e. the need to calibrate d-terms) into the synthetic observation with the ngEHT coverage. {We add d-terms randomly at all sites with a an error of roughly $5\%$ for $d_r$ and $d_l$ independently, and gains with a standard deviation of $10\%$.} First we performed a Stokes I reconstruction with MOEA/D. As described in \citetalias{Mueller2023}, this reconstruction uses the closure quantities only, hence is less prone to the gain uncertainties. We select the best image based on the accumulation point criterion and self-calibrate the data set with this Stokes I image. In the next step, we hold the total Stokes I structure constant and recover linear polarization. While a common reconstruction (and calibration) of total intensity and polarized structures is desired and realized for instance in state-of-the-art Bayesian frameworks \citep{Broderick2020, Tiede2022}, we skip this approach to avoid too high dimensionality due to the large number of multiobjective functionals that we would need to combine. We calculate an initial guess for the polarimetric structure by \textit{ehtim} with an unpenalized reconstruction, similar in philosophy to the initial guesses that are typically used for \textit{DoG-HiT} \citep{Mueller2022}. This initial guess is used to perform an a-priori d-term calibration and to initialize the population for MOEA/D. {We have used ehtim} for the final d-term calibration and {then} imaging with MOEA/D. 

\begin{figure*}
    \centering
    \includegraphics[width=1\textwidth]{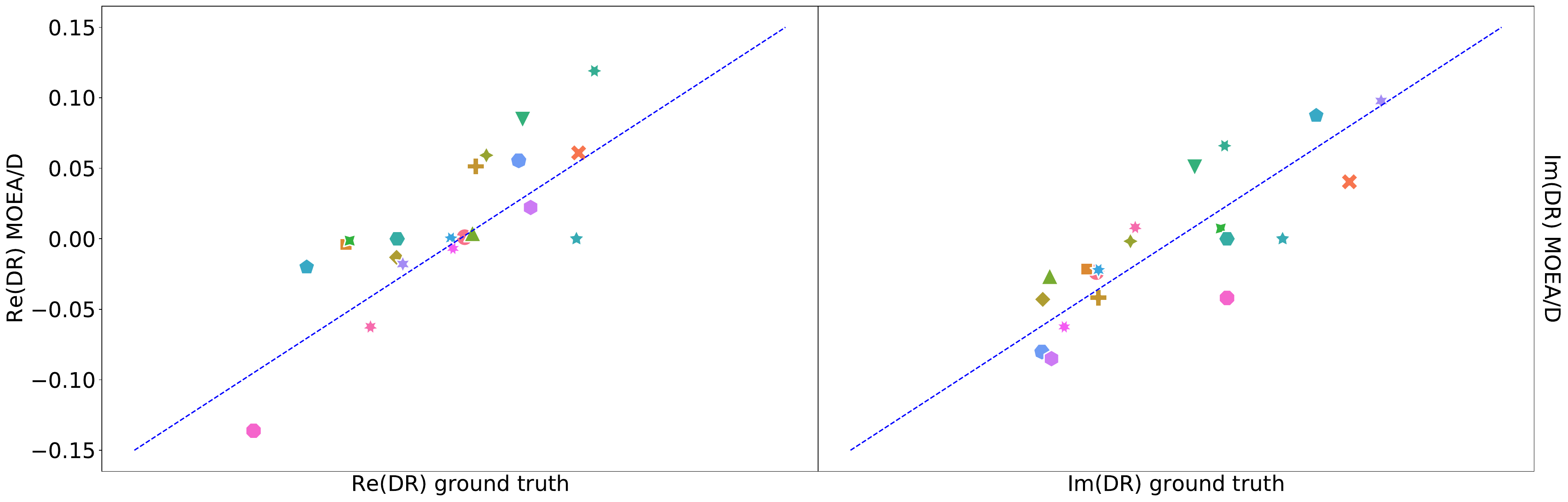}
    \includegraphics[width=1\textwidth]{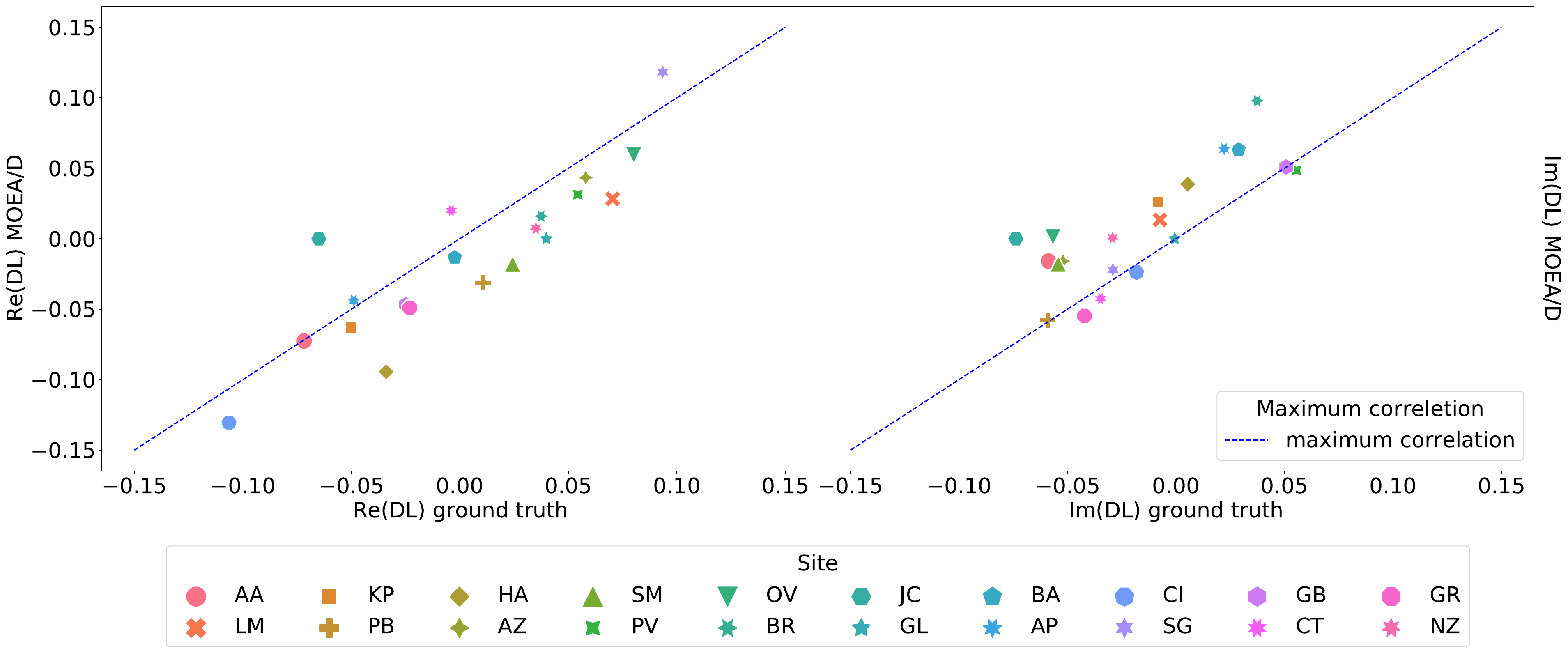}
    \caption{{D-terms comparison between the ground truth simulated and the recovered by MOEA/D using closure-only of the source showed in Fig.~\ref{fig: dcal_pol}. Different sites are represented by different colors and markers explained in the legend. The dashed blue lines indicates a perfect correlation. The closer the derms to the line the better correlation between the terms. \textit{Top row:} DR dterms (real and imaginary part from left to right). \textit{Bottom row:} DL dterms.}}
    \label{fig:dr_dl_comp}
\end{figure*}

{In Fig.~\ref{fig:dr_dl_comp} we show the correlation of MOEA/D obtained dterms and the ground truth. We can qualitatively see how we recover a close correlation (dashed blue line), implying that MOEA/D solution has dterms similar to the ground truth. In Appendix~\ref{app:dterms} the specific value of the dterm for both, true and solution, for every site can be found.}

\begin{figure*}
    \centering
    \includegraphics[width=\textwidth]{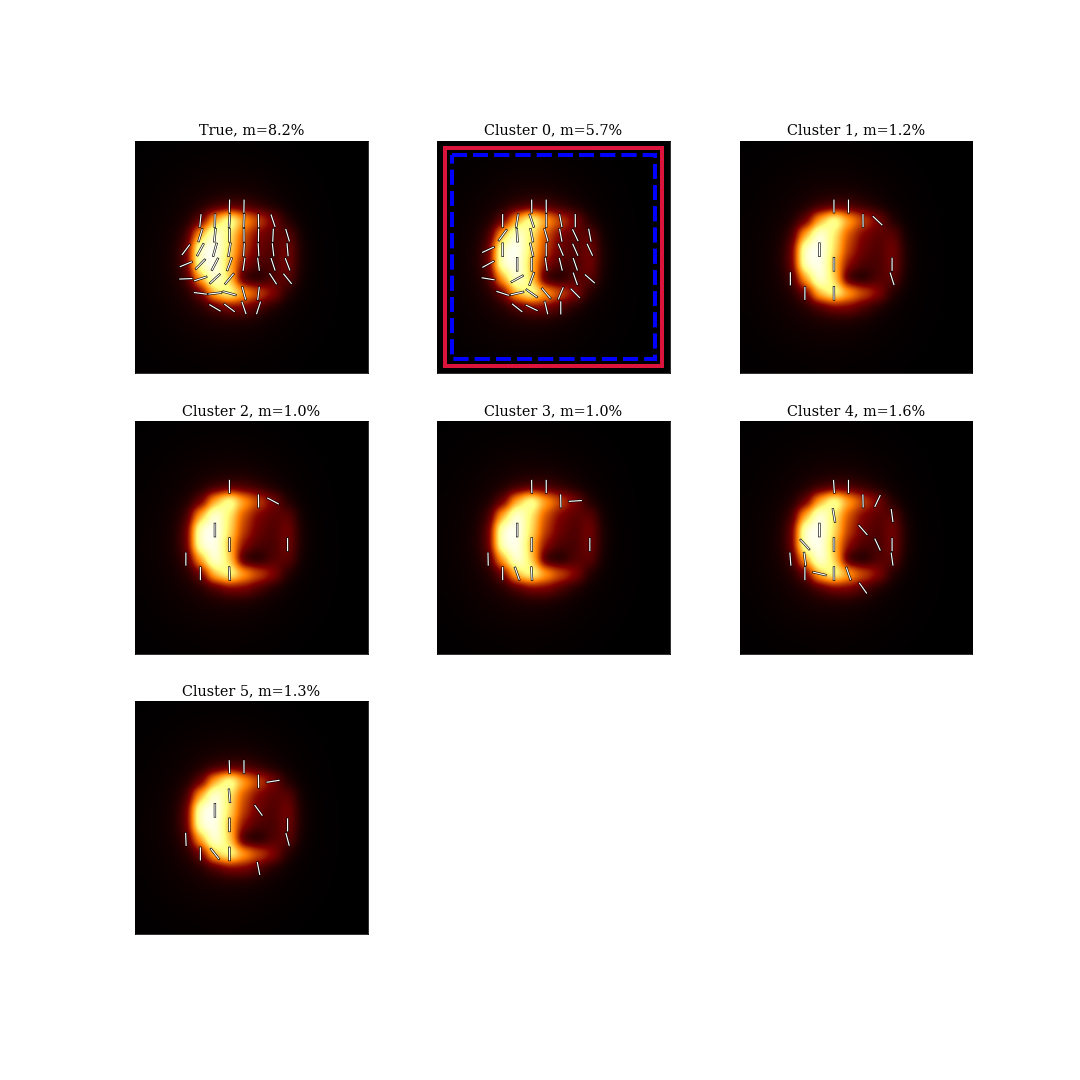}
    \caption{Same as in Fig. \ref{fig: pol_ngeht_sgra}, but for this synthetic data set we applied additional d-term errors.}
    \label{fig: dcal_pol}
\end{figure*}

MOEA/D approximates the overall structure during the first few iterations, timely evolving the front towards more fine-structure in later iterations. Hence, it is a natural approach, to use imaging and calibration in an alternating mode: We evolve the initial population by MOEA/D for some iterations, use the current best guess model to calibrate d-terms, and rerun MOEA/D with the old population as an initial guess (i.e. we let the population adapt to slightly varied objectives in an evolutionary way). We iterate this procedure three times. The corresponding reconstruction result is shown in Fig. \ref{fig: dcal_pol}. The overall structure of the Pareto front is similar to the fronts presented before in Fig. \ref{fig: pol_eht_sgra} and Fig. \ref{fig: pol_ngeht_sgra} with one dominating cluster. We recover the structure in the EVPAs successfully, however we would like to highlight that the linear polarization fraction gets underestimated. This may be a consequence of the self-calibration procedure since an initial guess with a smaller polarization fraction than the true one was used.

\subsection{Dynamics}
\label{subsec:mj_results}

The capability of recovering dynamic reconstructions using MOEA/D has been tested using two independent synthetic data simulating a SMBH with an accretion disk based on General Relativity Magneto Hydrodynamics (GRMHD) models.
The first data set, a more simple one, is based on the one presented in~\cite{Johnson2017, shiokawa2013}, and the data was shared by private communication. This data only contains Stokes I information. We call this data \texttt{MJ2017}.
{In similarity with Appendix A~\citetalias{Mueller2023}, we have performed a small survey to find values that better show numerical performance for the $\mu_{\mathrm{ngMEM}}$ initial value. We have found that one is a good starting value.}

The second synthetic data, called \texttt{CH3} set is the one given for the \texttt{Challlenge 3} presented in~\cite{ngehtchallenge}\footnote{\url{https://challenge.ngeht.org/}}. The synthetic data of this last dataset is dynamically polarized, i.e., the intrinsic polarization structure of the source varies along the observation. Therefore, we show an example of a joint polarimetric and dynamic reconstruction, and thus we will see how MOEA/D solves for polarimetric movies and dynamics are not limited to Stokes I 

To assess the performance of MOEA/D in dynamic reconstruction, we replicated the experiment conducted by~\cite{Johnson2017}, based on a Gerneral Relativistic Magento-Hydrodynamic (GRMHD) simulation of a black hole. More details can be found in the cited paper and in~\cite{shiokawa2013}. Specifically, we reconstructed a similar keyframes presented in Figure 4 of their work, which correspond to time instances at 16:00 UTC, 00:43:26 UTC and 06:29:27 UTC. These keyframes represent subsets of antennas observed within the EHT 2017 array configuration. The first keyframe involves $\sim 4$ antennas, the second $\sim 5$ antennas, and the final one only $\sim 2$ antennas. The differing numbers of baselines between these keyframes render the reconstruction of a continuous movie through snapshot imaging unfeasible. A comprehensive comparison between the performance of ngMEM and the regularizers proposed by~\cite{Johnson2017} in this specific scenario is detailed in~\cite{Mus2023}.

To generate a comparable movie, we initially construct a tentative movie using the constrained ngMEM. Subsequently, we flagged all visibilities corresponding to the aforementioned timestamps. Employing the ``guess movie'' as an initial point of our algorithm, we executed MOEA/D to explore the neighborhood of the movie.

\begin{figure*}
    \centering
    \includegraphics[width=0.5\linewidth]{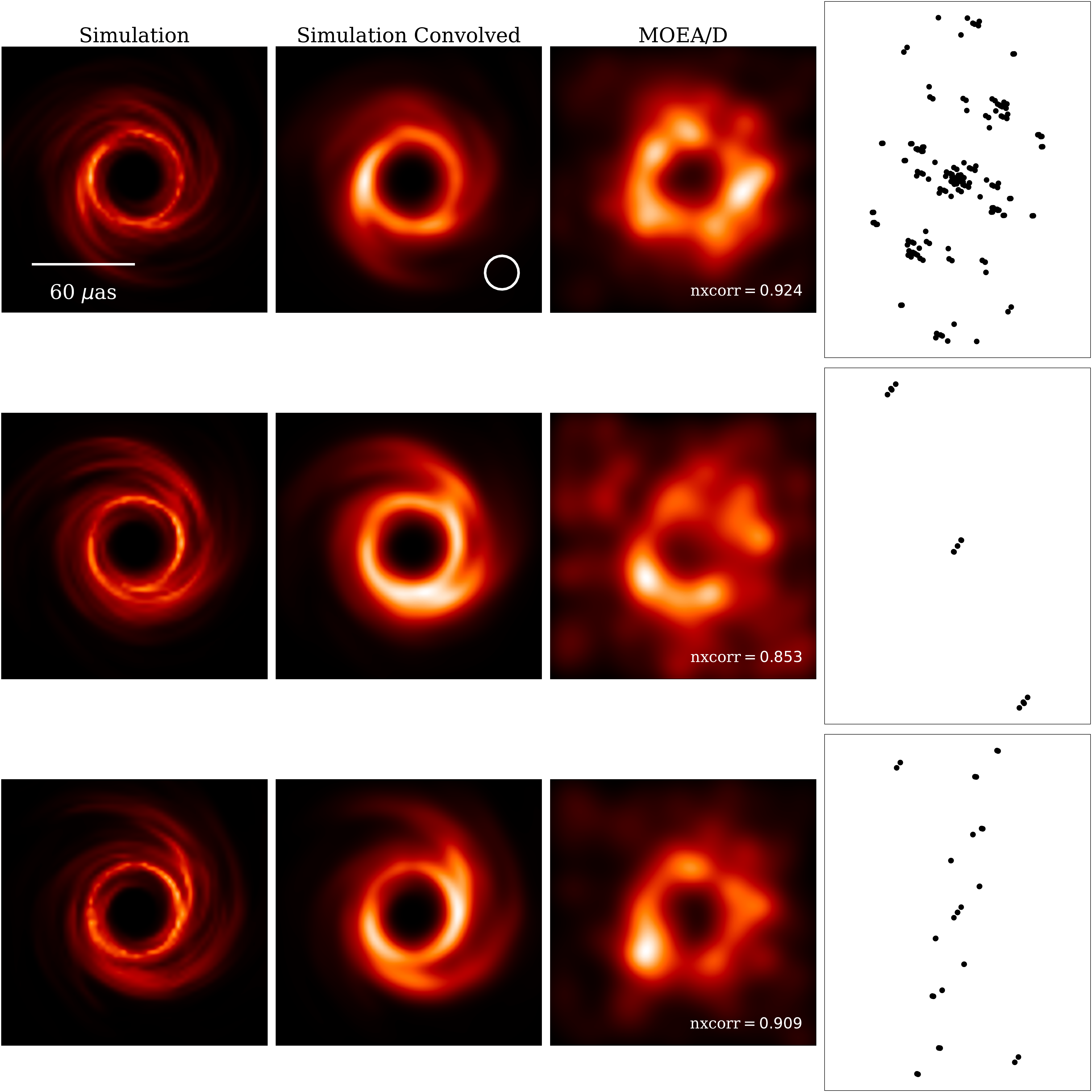}
    \caption{Comparison reconstructions of~\citet{Johnson2017} method and MOEA/D for the different uv-coverages of each keyframe. The color scale is linear and is consistent among different times, but is scaled separately for each case, based on the maximum brightness over all frames. The normal cross-corrleation (\texttt{nxcorr}) value between the recovered and the convolved model can be found in the bottom corner of the reconstructed images. {A frame-wise comparison can be found in~\cite{MusThesis}.}}
    \label{fig:mj_reconstruction_case}
\end{figure*}

We present our reconstruction results in Fig.~\ref{fig:mj_reconstruction_case}. Despite the extreme sparsity of the observation, MOEA/D successfully captures the fine crescent structure for all three snapshots. First column of the figure represents the simulation with infinite resolution. The middle column, the simulation convolved with the beam of the EHT $\sim 20\mu\mathrm{arcsec}$ (indicated as a with circle), and the third column, one representative solution of the MOEA/D convolved with the same beam. In Fig.~\ref{fig:pareto_mj} of Appendix~\ref{app:pareto_mj}, the Pareto front for each of the keyframes is presented.

The obtained results exhibit a performance deficit compared to those demonstrated in~\cite{Mus2023}. This discrepancy can be attributed to the inherent loss of cross-talk information between frames in our methodology. The cited paper use a Hessian modeling. Hessian encodes correlation information of the frames. Therefore, a better reconstruction can be done. However, it is important to acknowledge that this choice restricts the optimization process to a local scope, potentially resulting in the loss of the {possibly} inherent multimodal nature of the problem.

\subsection{Dynamic Polarimetry}
\begin{figure*}
    \centering
    \includegraphics[width=\textwidth]{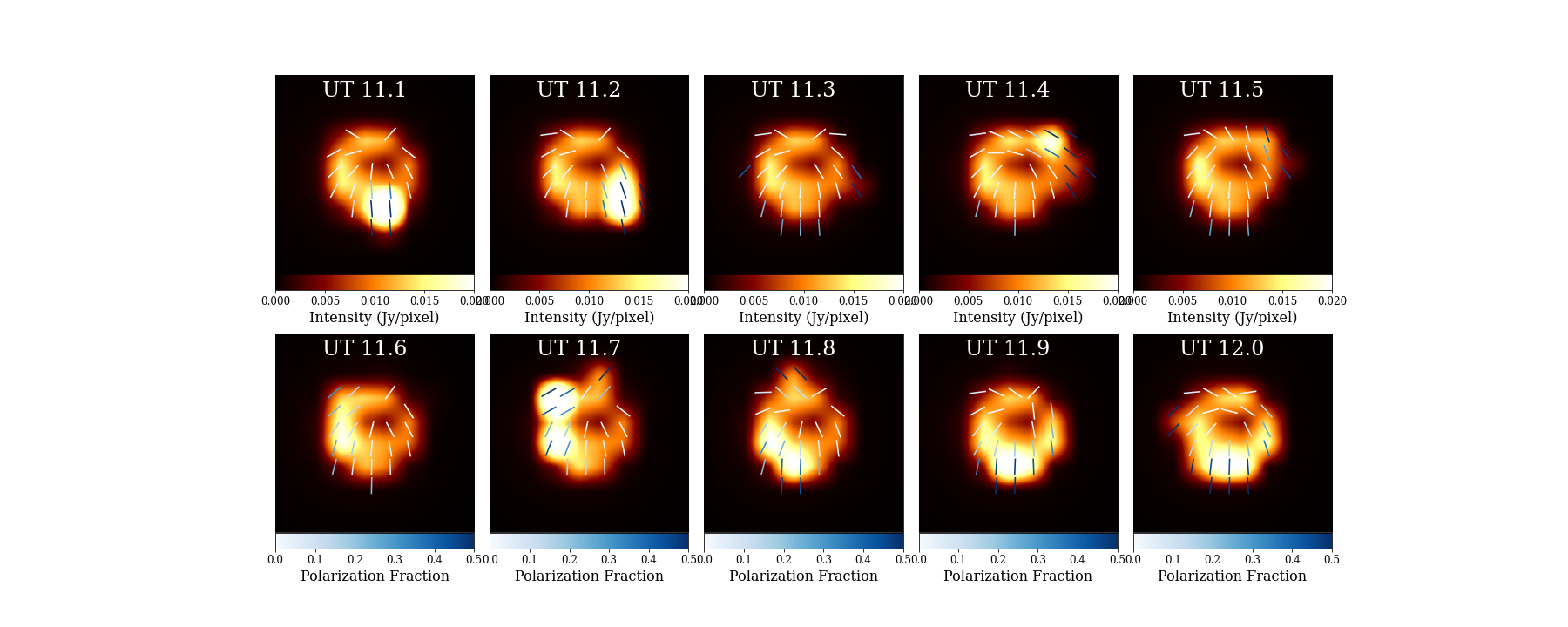}
    \caption{True movie of the ngEHT Challenge 3 synthetic data~\citep{ngehtchallenge}. Observation has been divided in 10 keyframe, each of them has been regridded to the MOEA/D resolution, i.e. to $10\,\mu\mathrm{as}$ pixels. EVPA are represented as lines and their corresponding polarization fraction appears as color map: the bluer, the stronger $m$. The intensity of the source is represented by other color map.}
    \label{fig: true_movie}
\end{figure*}

We test the capability to do dynamic polarimetry with a synthetic data set that is based on the ngEHT Analysis challenges \citep{ngehtchallenge}. Particularly we use the SGRA\_RIAFSPOT model from the third {ngEHT Analysis} challenge \footnote{\url{https://challenge.ngeht.org/challenge3/}}. The ground truth movie of Sgr A* is a RIAF model \citep{Broderick2016} with a shearing hotspot \citep{Tiede2020b} inspired by the observations of \citet{Gravity2018}. For more details on the simulation we refer to \citet{ngehtchallenge, Chatterjee2023}. We show the ground truth movie in Fig. \ref{fig: true_movie}. For the plotting, we regridded every frame to the field of view and pixel size that was adapted for MOEA/D, i.e. we represent the image structure by $10\,\mu\mathrm{as}$ pixels. We observed the movie synthetically by the possible EHT+ngEHT configuration specified in \citet{ngehtchallenge}. This configuration consists of all current EHT antennas and ten additional antennas from the list of proposed sites by \citet{Raymond2021}. We synthetically observe the source in a cadence with scans of ten minutes of observation, and two minutes gaps {between integration times}, as was done already in \citet{Mueller2022c}. The respective uv-coverage for the whole observation is presented in Fig. \ref{fig:uv_cov}. We add thermal noise, but assume a proper gain and d-term calibration. 

\begin{figure*}
    \centering
    \includegraphics[width=\textwidth]{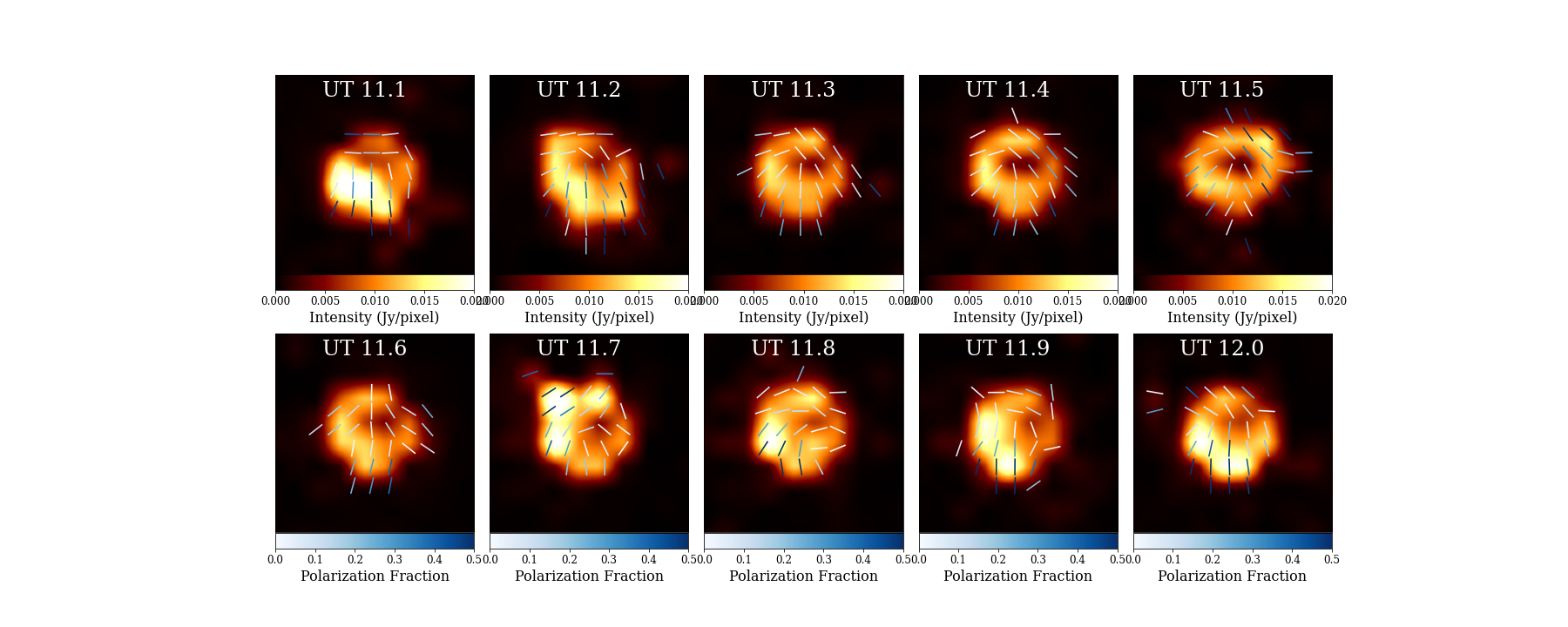}
    \caption{Recovered movie of the ngEHT Challenge 3 synthetic data~\citep{ngehtchallenge}. Each keyframe (which corresponds to exactly the same keyframes depicted in~\ref{fig: true_movie}) has been regridded to the MOEA/D resolution, i.e. to $10\,\mu\mathrm{as}$ pixels.}
    \label{fig: reco_movie}
\end{figure*}

\begin{figure*}
    \centering
    \includegraphics[width=\textwidth]{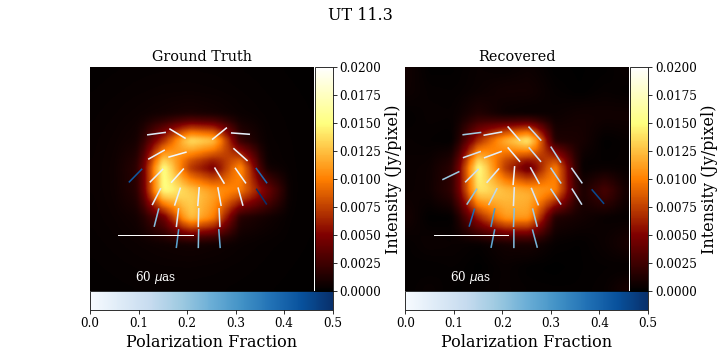}
    \caption{Single keyframe at UTC $11.3$ recovered from the third ngEHT Analysis Challenge, with the ground truth image on the left and the recovered keyframe in the right panel. The keyframe has been regridded to the MOEA/D resolution, i.e. to $10\,\mu\mathrm{as}$ pixels.}
    \label{fig: 11_3}
\end{figure*}

\begin{figure*}
    \centering
    \includegraphics[width=\textwidth]{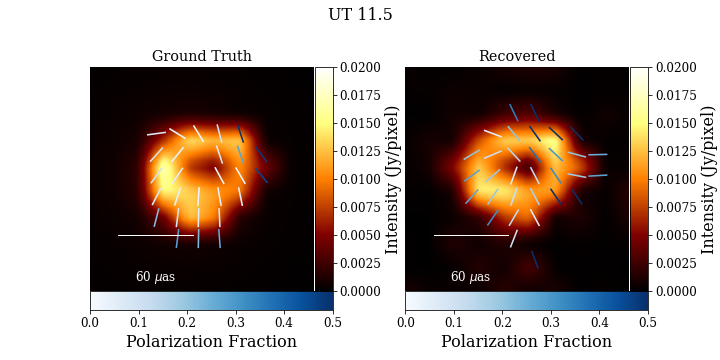}
    \caption{Same as Fig. \ref{fig: 11_3}, but at UTC $11.5$.}
    \label{fig: 11_5}
\end{figure*}

\begin{figure*}
    \centering
    \includegraphics[width=\textwidth]{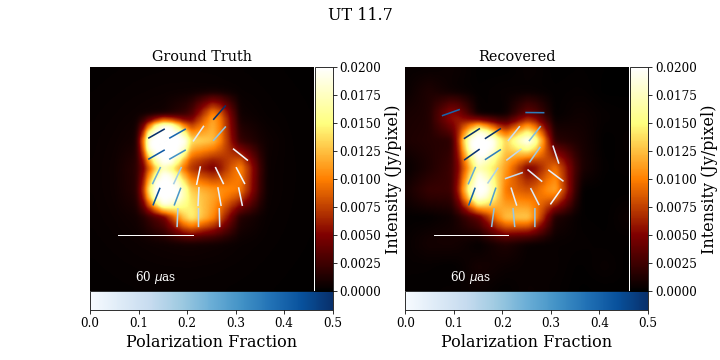}
    \caption{Same as Fig. \ref{fig: 11_7}, but at UTC $11.7$.}
    \label{fig: 11_7}
\end{figure*}

We show the recovered reconstruction in Fig. \ref{fig: reco_movie}. The polarimetric movie was recovered with snapshots of six minutes in the time-window with the best uv-coverage. For better comparison we show single snapshots in Figs.~\ref{fig: 11_3}, \ref{fig: 11_5} and \ref{fig: 11_7}. The ring-like image of the black hole shadow is successfully recovered at every keyframe with varying brightness asymmetry. Moreover, we successfully recover some hints for the dynamics at the event horizon in the form of a shearing hotspot, albeit the large pixel size limits the quality of the reconstruction. The overall EVPA pattern (orientation and strengths) is very well recovered in all keyframes. The dynamics of the EVPA pattern is well recovered as well as demonstrated for example by the well recovered enhanced polarization fraction towards the shearing hotspot in the top-right of Fig. \ref{fig: 11_5} and the top-left of Fig. \ref{fig: 11_7}.

\subsection{Quantitative evaluation}

To evaluate quantitatively the similarity of MOEA/D polarimetric dynamic reconstruction, we show to different figures-of-merits: the angular distribution of the reconstruction, the normal cross-correlation in total intensity ~\citep[see, for instance][]{Farah2022}), and $\beta_2$ to compare the EVPA pattern. First, is worth to notice that since ngMEM allows capture the evolution in every integration time, the comparison can be done in every frame (integration times).

\begin{figure}
    \centering
    \includegraphics[width=0.5\textwidth]{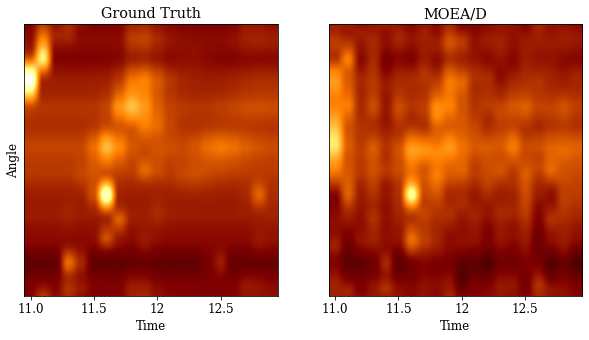}
    \caption{{Phase diagrams as presented in~\cite{Mus2023}. Horizontal axis represent the time (frames) and vertical axis the angle. Keyframes are linearly interpolated in every integration time. \textit{Left panel} shows the model correspondent to the dynamical movie. \textit{Right panel} is the recovered movie}. We can see how MMOEA/D is able to recover the hotpsot movement (the most bright lines).}
    \label{fig:profiles}
\end{figure}

\begin{figure}
    \centering
    \includegraphics[width=0.5\textwidth]{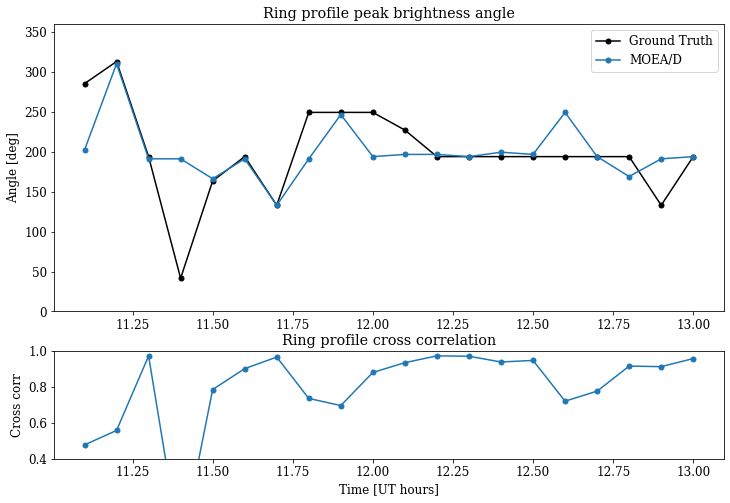}
    \caption{{\textit{Top panel:} Similar to~\ref{fig:profiles} but only the peak brightness angle versus time. Black line is the ground truth and blue line the recovered solution. \textit{Bottom panel:} \texttt{nxcorr} in every frame.}}
    \label{fig:pa}
\end{figure}

Regarding the Stokes I evolution, we present Fig.~\ref{fig:profiles},~\ref{fig:pa}. The first plot shows the angular brightness distribution in each integration time for both, model and reconstructed. The second plot compares the maximum bright peak of the phase of the ground truth and the model. This plot also shows the \texttt{nxcorr} frame-wise, being worst at the beginning of the experiment (as seen also with $\beta_2$) and improving during the observation.

\begin{figure}
    \centering
    \includegraphics[width=0.5\textwidth]{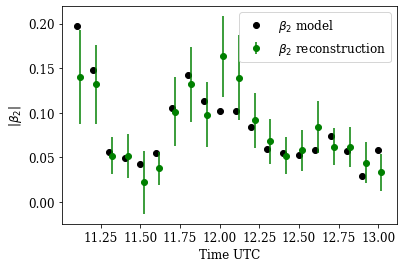}
    \caption{{Amplitude $\beta_2$ of the model (black dots), of the reconstruction (green dots) and the complex norm of their difference (blue dashed lines).}}
    \label{fig:beta2_comp}
\end{figure}

\begin{figure}
    \centering
    \includegraphics[width=0.5\textwidth]{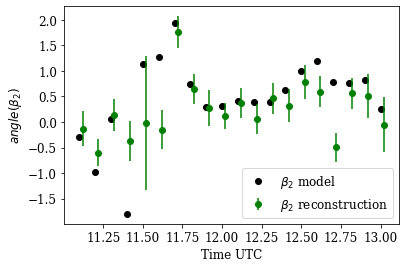}
    \caption{{As Fig.~\ref{fig:beta2_comp} but phases of $\beta_2$ are represented.}}
    \label{fig:beta2_angle}
\end{figure}

Fig.~\ref{fig:beta2_comp},~\ref{fig:beta2_angle} show the $\beta_2$ amplitude (first figure) and phase (second figure) evolution during the experiment. The parameter $\beta_2$ to describe the polarimetric signature in the EHT observations was introduced by \citet{Palumbo2020}. Because the use of closures during the optimization process, the absolute position of the source is lost. Therefore, in order to compare with the real source, we have aligned each keyframe of the reconstruction with the model by shifting the baricenter of the image to the North, East, South and West by one pixel. Errors bars in the $\beta_2$ reconstruction are the standard deviations by these shifts, representing the possible errors obtained by image alignment. In both, amplitude and phases we see a similar and congruent trend between the true solution and the recovered solution. However, the first keyframes recover worst $\beta_2$ in consistency with the findings in total intensity. Note that the first scans are also the scans in which the source is evolving the fastest, and thus the scans are more challenging recover.

\begin{figure}
    \centering
    \includegraphics[width=0.5\textwidth]{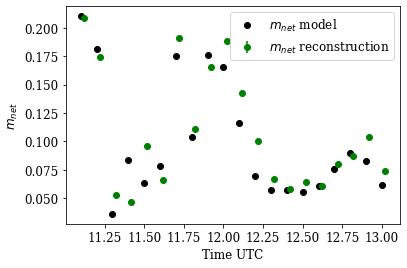}
    \caption{{The net polarization as a function of time.}}
    \label{fig:mnet}
\end{figure}

{We show in Fig. \ref{fig:mnet} the net polarization~\citep[for instance][]{EHTVII} evolving in time. MOEA/D performs quite well in recovering the correct polarization fraction at all times, even in the rapidly evolving first part of the evolution at which MOEA/D had more issues with finding the twistiness of the pattern (as indicated by $\beta_2$). MOEA/D's potential to recover the correct polarization fraction with the EHT+ngEHT coverage has already bee shown in Fig. \ref{fig: pol_ngeht_sgra}. It is noteworthy that while we spotted an underestimated polarization fraction for leakage-corrupted data, the time-dynamic (but fully calibrated) reconstruction with the same array does not show this tendency.}

\section{Summary and Conclusions}
\label{sec:summary}

Imaging reconstruction of interferometric data is a challenging ill-posed problem, particularly when dealing with very sparse uv-coverage, as is often the case in global VLBI observations. These challenges are amplified when working with polarimetric data or attempting to capture the dynamic behavior of the source, and become even more complex when reconstructing evolving polarimetric data.

On the one hand, while static polarimetry imaging has been extensively studied in the past~\citep{Ponsonby1973,Narayan1986,Holdaway1988,Coughlan2013,Chael2016}, solving the problem of polarimetric {multiobjective} imaging remains an open challenge.

On the other hand, the intricacies presented by rapidly evolving sources, such as the case of \sgra~\citep{eht2022c}, have prompted the VLBI community to develop innovative algorithms capable of effectively addressing variability concerns. The inherent limitations of snapshot imaging due to restricted coverage and the potential loss of information resulting from temporal averaging highlight the necessity for the formulation of functionals that can efficiently mitigate such information loss.

In this study, we have extended the MOEA/D algorithm, which was initially introduced in~\citepalias{Mueller2023}, to tackle the challenges associated with imaging Stokes I, Q, U parameters, fractional polarization ($m$), and dynamics. This enhanced iteration of MOEA/D showcases its efficiency in reconstructing both static and dynamic interferometric data.

In this work, we first introduced the modelization of MOEA/D for static polarimetry and dynamic Stokes I by incorporating the relevant functionals~\eqref{eq: pol1},~\eqref{eq: pol2},~\eqref{eq: pol3},~\eqref{eq: pol4} (for polarimetry) and ngMEM~\citep{Mus2023}~\eqref{eq: dyn7} for dynamic recovery. Subsequently, we tested our algorithm using synthetic data for both scenarios: polarimetry and dynamics. Finally, utilizing synthetic data from EHT+ngEHT, we demonstrated how MOEA/D excels in recovering polarimetric sequences, a capability possessed by only a select few algorithms.

The main benefits of MOEA/D for VLBI imaging are as follows. Opposed to classical RML and Bayesian techniques MOEA/D self-consistently {explores imaging problem globally with multiple image modes}, i.e. the issue of missing data for sparsely sampled VLBI observations. In this way, MOEA/D presents a robust, alternative claim of the image structure. Moreover, due to the full exploration of the Pareto front, parameter surveys are not needed to establish the estimate of the image, thus MOEA/D is a significant step towards an effectively unsupervised imaging procedure.
It is noteworthy that MOEA/D handle the non-negativity by imposing bounds on the image structure during the genetic operations rather than using the lognormal transform for the functional evaluations or adding lower-bound-constraints. {In the same way, $m$ is constrained to be in $\left[0,1\right]$ during the genetic evolution process. However, solutions accumulating at the edges of this interval are common due to the polarimetric entropy, and the random nature of the genetic algorithm.}

In conclusion, in the set of these two papers, we have shown how MOEA/D is able to recover static and dynamic imaging algorithm, for Stokes I and polarimetry, effectively mitigating some of the limitations associated with the current state-of-the-art RML methods. Its flexibility and proficiency in recovering Pareto optimal solutions render this algorithm an invaluable tool for imaging reconstruction of interferometric data, catering to both current and next-generation telescopes.

One remaining weakness of MOEA/D is its current limitation to a small number of pixels (256 in this set of two papers). To overcome this problem, an alternative formulation is done in~\cite{Mus2023b} in which the evolutionary search algorithm is only used to find the optimal combination of weights. The authors develop a novel strategy based on swarm intelligence to find the optimal weights associated to the RML problem, regardless the quality of the obtained image. Once the optimal weights are found, the corresponding image is, by definition, a member of the Pareto front, and thus, a valid and desirable solution. Then, an optimal image can be found using a local search algorithm, as L-BFG-S~\citep{LBFGS} imposing a correlation between close pixels (via a quasi-Hessian). The resulting image is so-called ``Shapley reconstruction''.

\section*{Software Availability}
We will make our imaging pipeline and our software available soon in the second release of MrBeam \footnote{\url{https://github.com/hmuellergoe/mrbeam}}. Our software makes use of the publicly available ehtim \citep{Chael2016, Chael2018}, regpy \citep{regpy}, MrBeam \citep{Mueller2022, Mueller2022b, Mueller2022c} and pygmo \citep{Biscani2020} packages.

\begin{acknowledgement}
AM and HM have contributed equally to this work. This work was partially supported by the M2FINDERS project funded by the European Research Council (ERC) under the European Union’s Horizon 2020 Research and Innovation Programme (Grant Agreement No. 101018682) and by the MICINN Research Project PID2019-108995GB-C22. AM and IM also thanks the Generalitat Valenciana for funding, in the frame of the GenT Project CIDEGENT/2018/021 and the Max Plank Institut für Radioastronomie for covering the visit to Bonn which has made this possible. HM received financial support for this research from the International Max Planck Research School (IMPRS) for Astronomy and Astrophysics at the Universities of Bonn and Cologne. Authors acknowledge Michael Johnson for sharing via private communication their synthetic data.
\end{acknowledgement}

\bibliographystyle{aa}
\bibliography{lib}{}

\appendix

\section{Pareto fronts of the \texttt{MJ2017} data}
\label{app:pareto_mj}

In this Appendix, we present the three Pareto fronts, each corresponding to a keyframe, concerning the \texttt{MJ2017} dataset detailed in Sect.~\ref{subsec:mj_results}.
\begin{figure*}
    \includegraphics[width=.5\textwidth]{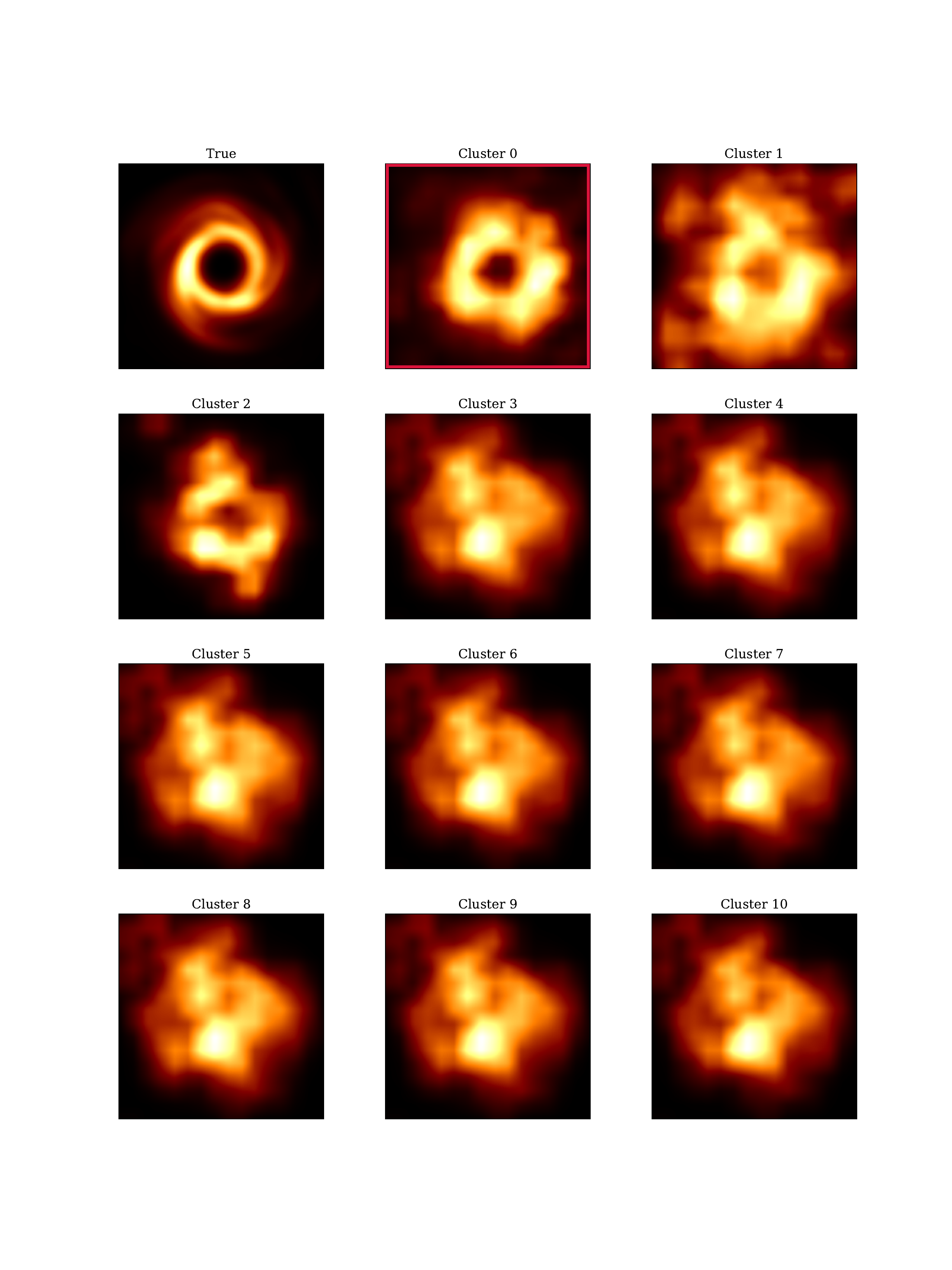}
    \includegraphics[width=.5\textwidth]{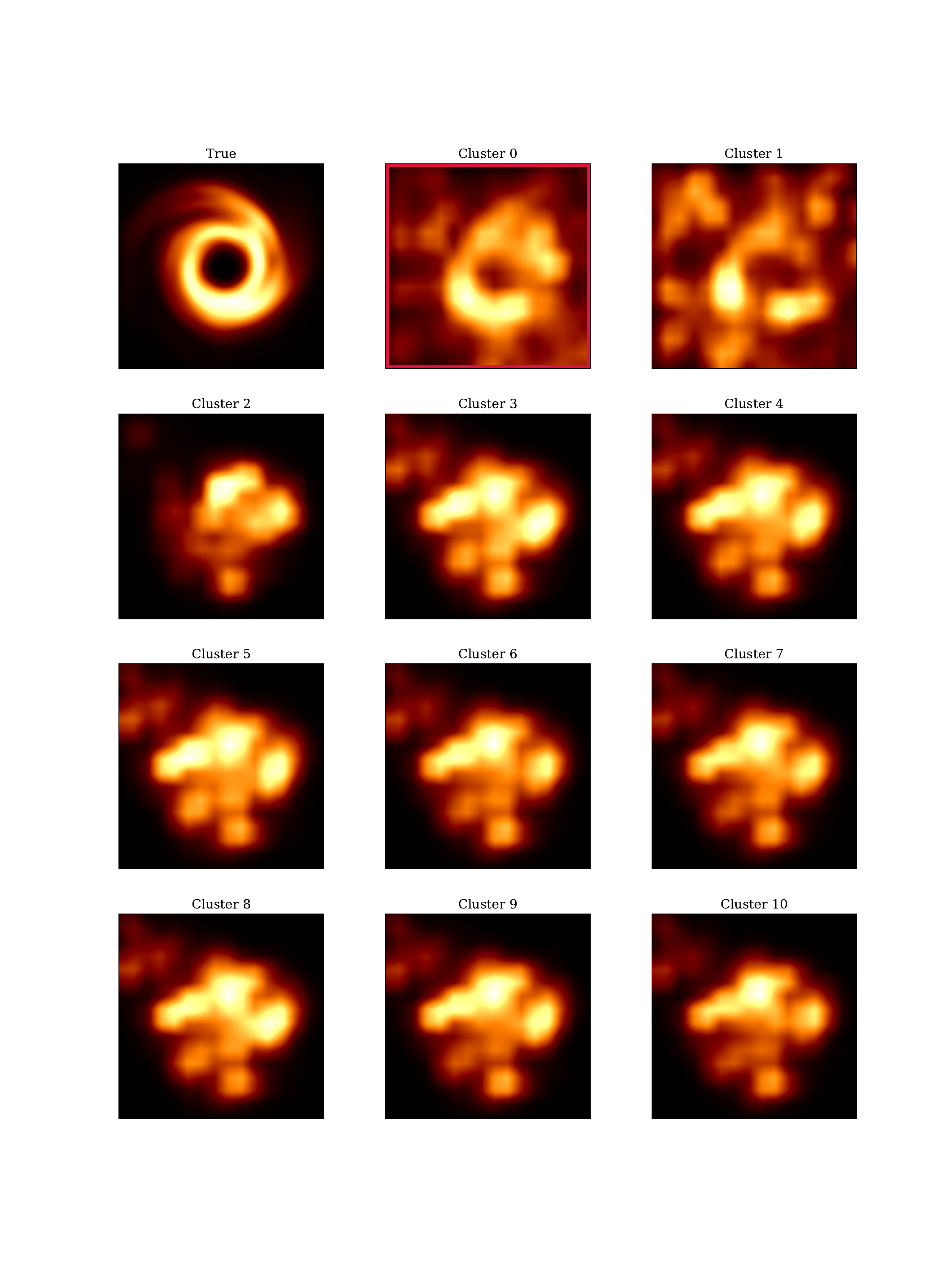}
    \\
    \includegraphics[width=.5\textwidth]{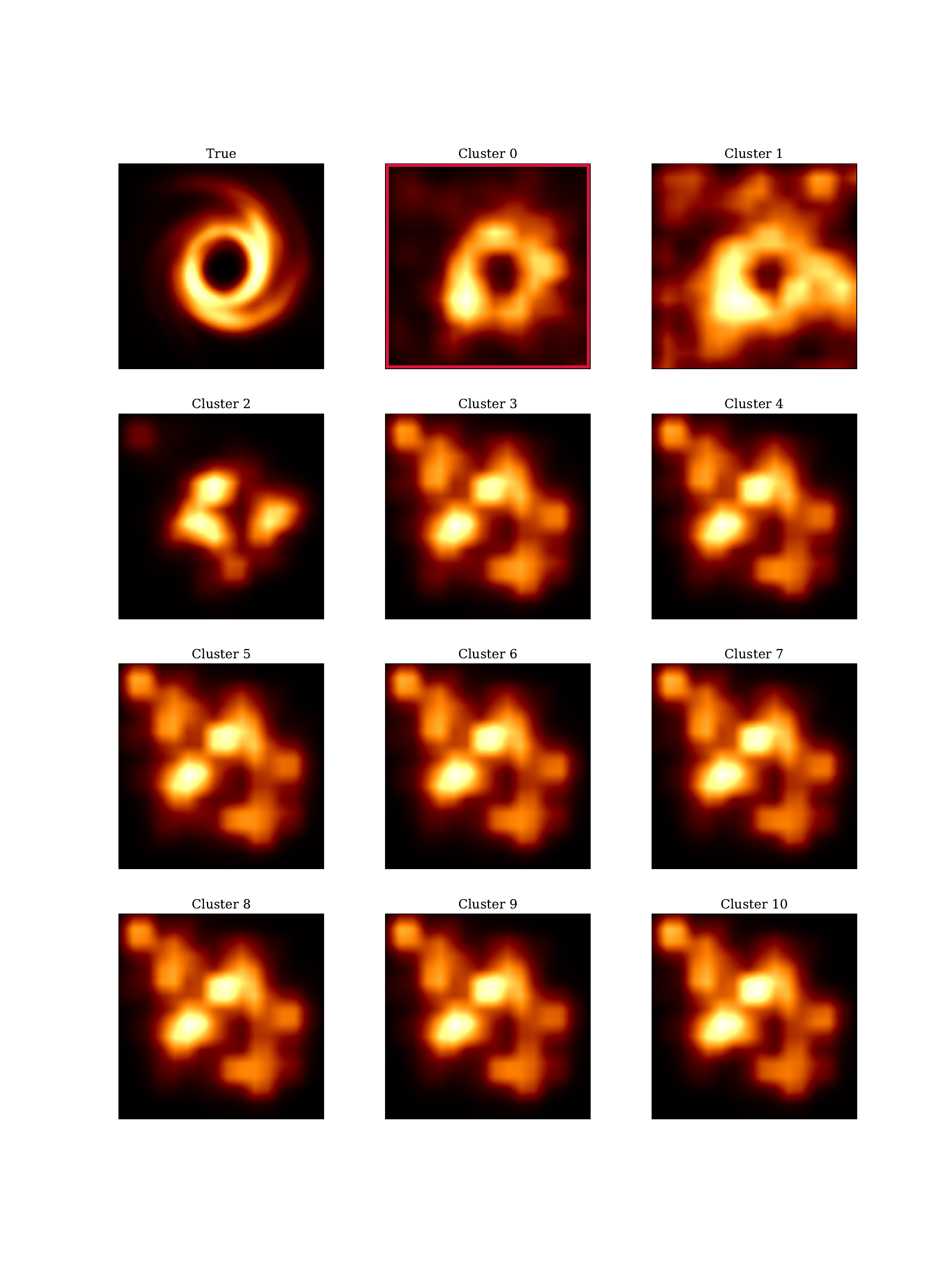}
    \includegraphics[width=.5\textwidth]{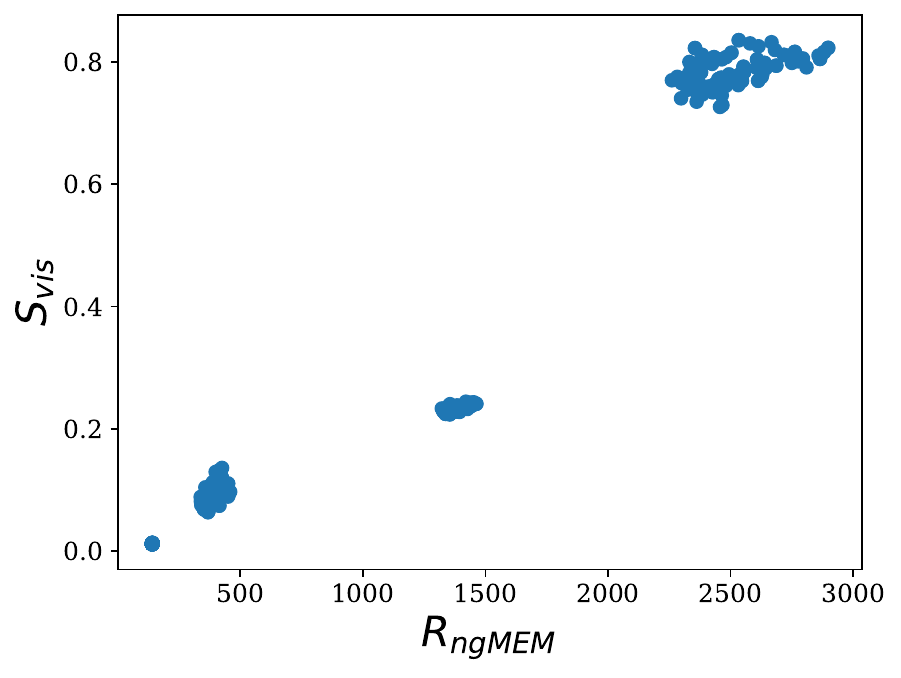}
    \caption{Pareto fronts for the first (top left), second (top right) and third (bottom left) keyframe for the \texttt{MJ2017} data. {Bottom right panel shows the $\mu_{\mathrm{ngMEM}}$ dependence with respect to the data fidelity term.}}
    \label{fig:pareto_mj}
\end{figure*}

The keyframe solutions shown in Fig.~\ref{fig:mj_reconstruction_case}, are the favored cluster (indicated by the red box) of the three Pareto fronts.

The sparsity observed in the uv-coverage across each snapshot is responsible for the significant diversity observed in the solutions. Notably, it is interesting to observe that, in all three cases, only a limited number of clusters exhibit a ring-like structure. Nevertheless, it is worth highlighting that ring structures are the predominant and highly probable solutions. In fact, all of the selected solutions, particularly the most recurrent ones, conform to ring-like geometry.

{Lastly, from the bottom right panel, we can deduce that better movie reconstructions require higher weight for the time entropy, as expected.}

\section{Dterm values}
\label{app:dterms}
The specific dterms values of Fig.~\ref{fig:dr_dl_comp} are 

\begin{table*}
    \centering
    \caption{Dterms for the MOEA/D solution and ground truth.}
\begin{tabular}{llrrrr}
\toprule
Site &                  DR MOEAD &                  DL MOEAD &                  DR model &                  DL model \\
\midrule
AA &  0.001116-0.023605j & -0.072563-0.015891j &  0.002200-0.036700j & -0.071800-0.058800j \\
LM &  0.060947+0.040446j &  0.028275+0.013363j &  0.054700+0.079900j &  0.070400-0.007400j \\
KP & -0.003959-0.021501j & -0.063051+0.026047j & -0.052500-0.041100j & -0.050100-0.008200j \\
PB &  0.051365-0.041802j & -0.031139-0.057868j &  0.007300-0.035800j &  0.010700-0.059200j \\
HA & -0.013171-0.042990j & -0.094325+0.038679j & -0.029200-0.061400j & -0.034000+0.005400j \\
AZ &  0.059227-0.001727j &  0.043303-0.015965j &  0.012200-0.021000j &  0.058000-0.052100j \\
SM &  0.003936-0.026483j & -0.017905-0.017813j &  0.005800-0.058200j &  0.024300-0.054300j \\
PV & -0.001330+0.007314j &  0.031320+0.048570j & -0.050800+0.020600j &  0.054300+0.055500j \\
OV &  0.084536+0.050740j &  0.059367+0.001148j &  0.028900+0.008600j &  0.080100-0.056700j \\
BR &  0.119096+0.065994j &  0.016005+0.097833j &  0.062000+0.022400j &  0.037400+0.037400j \\
JC &  0.000000+0.000000j &  0.000000+0.000000j & -0.028900+0.023500j & -0.065000-0.073800j \\
GL &  0.000000+0.000000j &  0.000000+0.000000j &  0.053700+0.049100j &  0.039800-0.000600j \\
BA & -0.020075+0.087414j & -0.013271+0.063254j & -0.070600+0.064600j & -0.002400+0.028800j \\
AP &  0.000359-0.022044j & -0.043728+0.063789j & -0.004100-0.035700j & -0.048900+0.022200j \\
CI &  0.055516-0.080114j & -0.130665-0.023834j &  0.027100-0.061800j & -0.106400-0.018200j \\
SG & -0.017905+0.097750j &  0.118187-0.021940j & -0.026300+0.094500j &  0.093400-0.029100j \\
GB &  0.022197-0.085032j & -0.046719+0.050880j &  0.032600-0.057400j & -0.024900+0.050700j \\
CT & -0.006828-0.062479j &  0.019986-0.042466j & -0.003200-0.051500j & -0.004000-0.034700j \\
GR & -0.136182-0.041948j & -0.048996-0.054741j & -0.095100+0.023500j & -0.023100-0.042200j \\
NZ & -0.062465+0.008080j &  0.007345+0.000625j & -0.041200-0.018800j &  0.035000-0.029200j \\
\bottomrule
\end{tabular}
\label{tab:drdl_obt}
\end{table*}

\end{document}